\documentclass[preprint,graphicx]{aastex}

\lefthead{Smith et al.}
\righthead{The $u'g'r'i'z'$ Standard Star System}

\begin{document}

\title{The $u'g'r'i'z'$ Standard Star System}

\author{J. Allyn Smith\altaffilmark{1,2,\ref{Michigan},\ref{Wyoming}},
Douglas L. Tucker\altaffilmark{1,\ref{Fermilab}},
Stephen Kent\altaffilmark{\ref{Fermilab}},
Michael W. Richmond\altaffilmark{\ref{Rochester}},
Masataka Fukugita\altaffilmark{\ref{CosmicRay},\ref{IAS}},
Takashi Ichikawa\altaffilmark{\ref{Tohoku}},
Shin--ichi Ichikawa\altaffilmark{\ref{NAOJ}},
Anders M. Jorgensen\altaffilmark{\ref{LANL}},
Alan Uomoto\altaffilmark{\ref{JHU}},
James E. Gunn\altaffilmark{\ref{Princeton}},
Masaru Hamabe\altaffilmark{\ref{JWU}},
Masaru Watanabe\altaffilmark{\ref{ISAS}},
Alin Tolea\altaffilmark{\ref{JHU}},
Arne Henden\altaffilmark{\ref{Flagstaff}},
James Annis\altaffilmark{\ref{Fermilab}},
Jeffrey R. Pier\altaffilmark{\ref{Flagstaff}},
Timothy A. McKay\altaffilmark{\ref{Michigan}},
Jon Brinkmann\altaffilmark{\ref{APO}},
Bing Chen\altaffilmark{\ref{JHU}},
Jon Holtzman\altaffilmark{\ref{NMSU}},
Kazuhiro Shimasaku\altaffilmark{\ref{UTokyo}},
Donald G. York\altaffilmark{\ref{Chicago},\ref{EFI}}
}

\altaffiltext{1}{Equal first authors.} 
\altaffiltext{2}{Visiting Astronomer, United States Naval Observatory.} 
\altaffiltext{3}{University of Michigan, 
            Department of Physics,
            500 East University, Ann Arbor, MI 48109
  \label{Michigan}}
\altaffiltext{4}{Current Address: University of Wyoming,
            Department of Physics \& Astronomy,
            P.O. Box 3905, 
            Laramie, WY 82071
  \label{Wyoming}}
\altaffiltext{5}{Fermi National Accelerator Laboratory, 
            P.O. Box 500, Batavia, IL 60510
  \label{Fermilab}}
\altaffiltext{6}{Physics Department,
            Rochester Institute of Technology,
            85 Lomb Memorial Drive,
            Rochester, NY 14623-5603
  \label{Rochester}}
\altaffiltext{7}{Institute for Cosmic Ray Research, 
            University of Tokyo, Midori, Tanashi, Tokyo 188-8502, Japan
  \label{CosmicRay}}
\altaffiltext{8}{Institute for Advanced Study, 
            Olden Lane, Princeton, NJ 08540
  \label{IAS}}
\altaffiltext{9}{Astronomical Institute, 
            Tohoku University,
            Aoba, Sendai 980-8578, Japan
  \label{Tohoku}}
\altaffiltext{10}{National Astronomical Observatory of Japan,
            Mitaka, Tokyo, 181-8588 Japan
  \label{NAOJ}}
\altaffiltext{11}{Los Alamos National Laboratory, NIS-4, D448,
            Los Alamos, NM 87545
  \label{LANL}}
\altaffiltext{12}{Department of Physics and Astronomy, 
            The Johns Hopkins University,
            3701 San Martin Drive, Baltimore, MD 21218, USA
  \label{JHU}}
\altaffiltext{13}{Princeton University Observatory, 
            Princeton, NJ 08544
  \label{Princeton}}
\altaffiltext{14}{Department of Mathematical and Physical Sciences,
            Japan Women's University,
            Mejirodai, Tokyo, 112-8681 Japan
  \label{JWU}}
\altaffiltext{15}{Institute of Space and Astronautical Science, 
             Sagamihara, Kanagawa 229-8510, Japan
  \label{ISAS}}
\altaffiltext{16}{U.S. Naval Observatory, Flagstaff Station,
            P.O. Box 1149,
            Flagstaff, AZ  86002-1149
  \label{Flagstaff}}
\altaffiltext{17}{Apache Point Observatory, 
            P.O. Box 59,
            Sunspot, NM 88349-0059
  \label{APO}}
\altaffiltext{18}{New Mexico State University,
            Dept. 4500, Box 30001,
            Las Cruces, NM 88003 
  \label{NMSU}}
\altaffiltext{19}{Department of Astronomy and Research 
            Center for the Early Universe, 
            School of Science, University of Tokyo, 
            Bunkyo-ku, Tokyo, 113-0033 Japan
  \label{UTokyo}}
\altaffiltext{20}{The University of Chicago,
            Department of Astronomy and Astrophysics,
            5640 S. Ellis Ave.,
            Chicago, IL 60637
  \label{Chicago}}
\altaffiltext{21}{The University of Chicago,
            Enrico Fermi Institute,
            5640 S. Ellis Ave.,
            Chicago, IL 60637
  \label{EFI}}

\begin{abstract}
We present the 158 standard stars that define the $u'g'r'i'z'$
photometric system.  These stars form the basis for the photometric
calibration of the Sloan Digital Sky Survey (SDSS).  The defining
instrument system and filters, the observing process, the reduction
techniques, and the software used to create the stellar network are
all described.  We briefly discuss the history of the star selection
process, the derivation of a set of transformation equations for the
$UBVR_{\rm c}I_{\rm c}$ system, and plans for future work.
\end{abstract}

\keywords{catalogs --- stars: fundamental parameters --- standards}

\section{Introduction}

We present the newly established standard star network for the
$u'g'r'i'z'$ filter system \citep[see][]{fuk96}.  This standard star 
network was developed at the U.S.  Naval Observatory, Flagstaff Station.  
These stars form the basis for the photometric calibration of the Sloan
Digital Sky Survey (SDSS).  The SDSS uses a 2.5-m telescope at Apache
Point Observatory (APO) to produce a five-band, photometrically
calibrated digital imaging survey of $\pi$ steradians (10,000 square
degrees) of the Northern Galactic Cap \citep{GCRS98,york00} as one of
its major data products.

It is not our purpose here to describe in detail the full end-to-end
process of calibrating the SDSS photometric data.  That is the topic of a 
future paper.  Here, we merely wish to present a self-contained description 
of the standard star network upon which the SDSS photometry is based.  We do 
note, however, that one of the targets of the SDSS is to achieve
a level of photometric uniformity and accuracy such that the
system-wide rms errors in the final SDSS photometric catalog will be
less than 0.02~mag in $r'$, 0.02~mag in $(r'-i')$ and $(g'-r')$, and
0.03~mag in $(u'-g')$ and $(i'-z')$, for objects bluer than an M0
dwarf.  To meet this target, internal goals were set for the accuracy
of the primary standard star system: the uncertainty in the mean
calibrated magnitudes for any given primary standard star should be
$\le$1.5\% at $u'$, $\le$ 1\% in $g', r'$ and $i'$, and $\le$1.5\% at
$z'$.  As we will show later in this paper, we more than meet these
goals for all but a handful of stars.

In addition, we must mention that, due to small-but-significant
differences between the USNO and 2.5-m filters, the final 2.5-m SDSS
published photometry will likely differ systematically from the
$u'g'r'i'z'$ system at the few percent level for $g'r'i'$ and slightly
worse for $u'$ and $z'$ \citep[see][]{sto01}.  When transformation
equations between the $u'g'r'i'z'$ system and the 2.5-m SDSS
photometry have been robustly determined, they will be promptly made
available to the astronomical community.  (Note: the intended accuracy
of these transformation equations is included within the
above-mentioned error budget for the photometric calibrations of the
final SDSS imaging catalog.)

The nomenclature used in this system differs slightly from the traditional 
photometric literature.  This was done to avoid confusion with existing
SDSS papers and nomenclature.  In the traditional sense, Vega ($\alpha$ Lyr) 
is the ultimate ``fundamental'' standard.  However, in this paper we refer 
to three subdwarf stars which were used to set the initial system zeropoint 
as ``fundamental'' with the other 155 stars of the system being referred
to as ``primary'' stars.  The term ``secondary'' is used within the SDSS
nomenclature to refer to the photometric system transfer patches --- pieces
of the sky that are observed by a 0.5-m telescope that are used to transfer
the photometric solution to the main survey imaging telescope. 

In the following sections we present details of the standard star
development program.  We describe the instrumentation and filter
system in \S2, selection of the initial set of stars in \S3, and a
brief overview of the reduction software that was used to develop the
network in \S4.  Final results for the inital set of $u'g'r'i'z'$
primary standard stars are presented in \S5, and we discuss future
extensions to this system in \S6.

\section{Observations and Instrumentation}

The observations  were obtained using  the 1.0-m Ritchey--Chr\'{e}tien
telescope at the  U.S. Naval Observatory's Flagstaff Station (USNO-FS)
during the   bright period of  each lunation  from  March 1998 through
January 2000  inclusive.  Since the program  stars  were fairly bright
the moon did   not unduly hamper   the observations;  however,  we did
maintain a minimum 30$\arcdeg$ radius of avoidance near the moon that,
for the most part, allowed us to use the dome  to shield the telescope
from direct moonlight.  This observing restriction had a slight impact
on the  choice of stars observed each  night.  The few ``faint'' stars
(fainter than about $r'$=13.2) on the list were observed when the moon
was below or within one hour of the opposite  horizon from the star so
the dome could be used to block the moon completely.

All of the observations were direct exposures  with a thinned, UV-AR
coated, Tektronix TK1024 CCD operating at a gain of 7.43$\pm$0.41
electrons per ADU with a readnoise  of 6.0 electrons.  This CCD is
similar to the CCDs used in the main SDSS  survey camera and the CCD
used by the 0.5-m photometric monitoring telescope at APO.  The camera
scale of 0.68 arcsec/pixel produced a  field of view of 11.54 arcmin.

\begin{figure}
\figurenum{1}
\includegraphics[angle=-90,scale=.75]{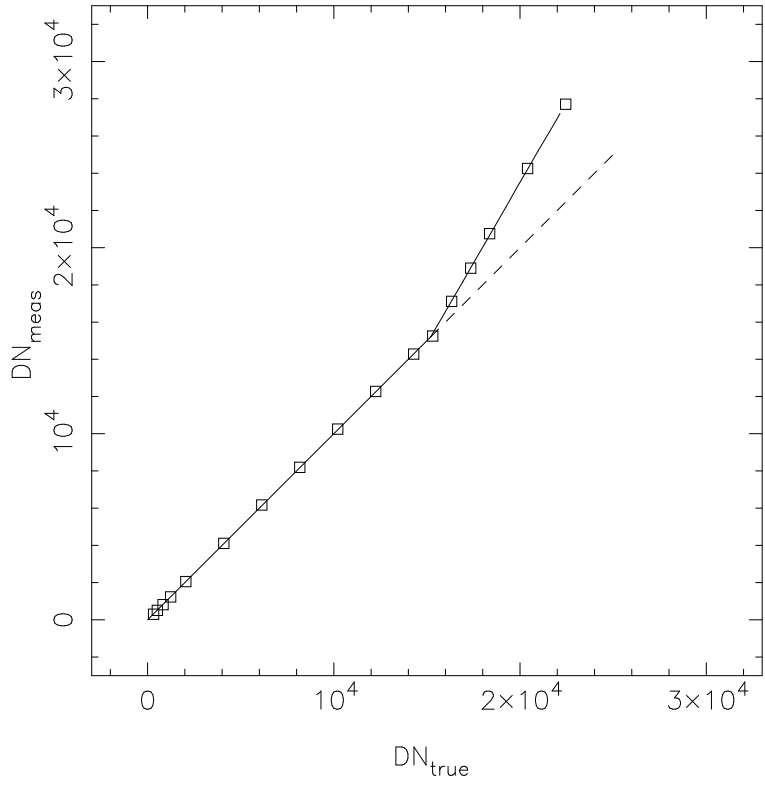}
\caption{The linearity curve for the TK1024 CCD used on the USNO 1.0-m 
telescope for this program (solid line).  $DN_{\rm meas}$ is the raw,
bias-subtracted value of the signal; $DN_{\rm true}$ is the value that
would have been measured if the CCD were completely linear.  Note the
``knee'' at $DN \approx 15,200$~ADU.  The dashed line acts as a
reference for what a fully linear relation would look like.
\label{usnoLinearity}}
\end{figure}

Testing of the chip revealed a linear response up to 15,200 ADU and a
second, well behaved and correctable linear response region up to
27,500 ADU (see Figure~\ref{usnoLinearity}).  This change in the CCD
response function, and sense of the change, is due to the clocking scheme
that was employed in the TK1024 read electronics.  The integration and
the lowering of the transfer gate occurred simultaneously.  The net
result is that, since the summing well has more charge for brighter
objects, the charge begins to `spill over' the gate potential faster for
these objects, resulting in an effective longer integration time (and a
seemingly higher gain) than for the faint objects. 

Exposure lengths were tailored to maximize the  number of potential standard 
stars in each field with good photon counts in all  five filters while not 
exceeding the first linear portion of the response curve for the primary star 
of interest.  Tailoring the exposure  lengths allowed us to develop multiple 
standards  in several fields.  Use of the linearity response curve allowed 
correction of the brighter stars when good atmospheric seeing caused counts 
greater than 15,200 ADU per pixel while allowing us to maximize  observing 
efficiency in some of the more populated fields.  The drawback of targetting 
multiple stars per field was a decrease in the signal to noise ratio for the 
extreme red or blue stars within the same field for which the exposures were 
not optimized.

\placefigure{usnoLinearity}

\begin{figure}
\figurenum{2}
\epsscale{0.75}
\includegraphics[angle=-90,scale=.75]{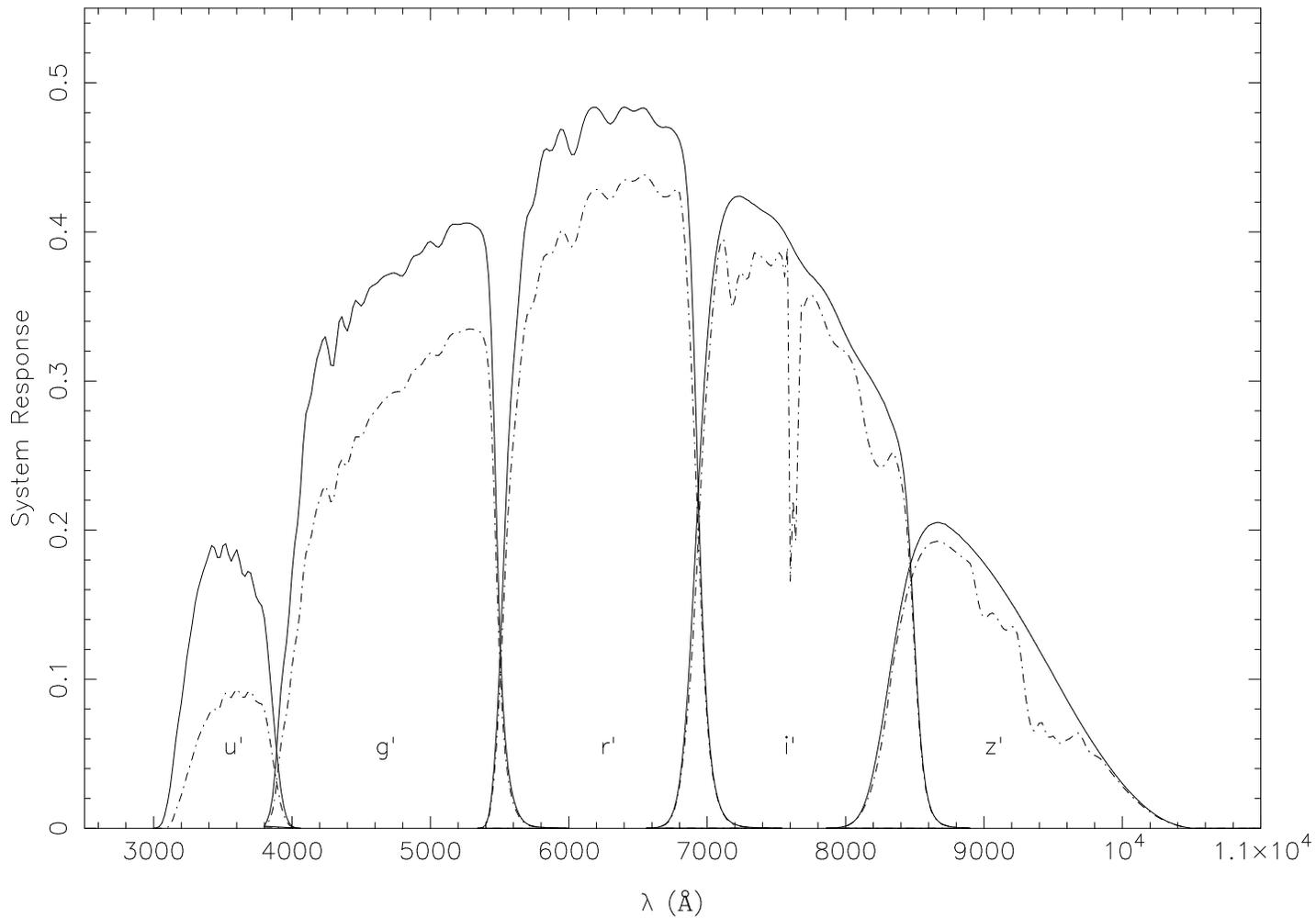}
\caption{The $u'g'r'i'z'$ system filter bandpasses convolved with a typical 
coated CCD. The curves represent the expected total quantum efficiencies of 
the camera plus telescope on the sky.  Solid curves indicate the response 
function without atmospheric extinction; dashed curves include extinction 
at 1.2 airmasses at the altitude of the U.S. Naval Observatory's Flagstaff 
Station.
\label{usnoResponse}}
\end{figure}

\begin{figure}
\figurenum{3}
\epsscale{0.75}
\includegraphics[angle=-90,scale=.75]{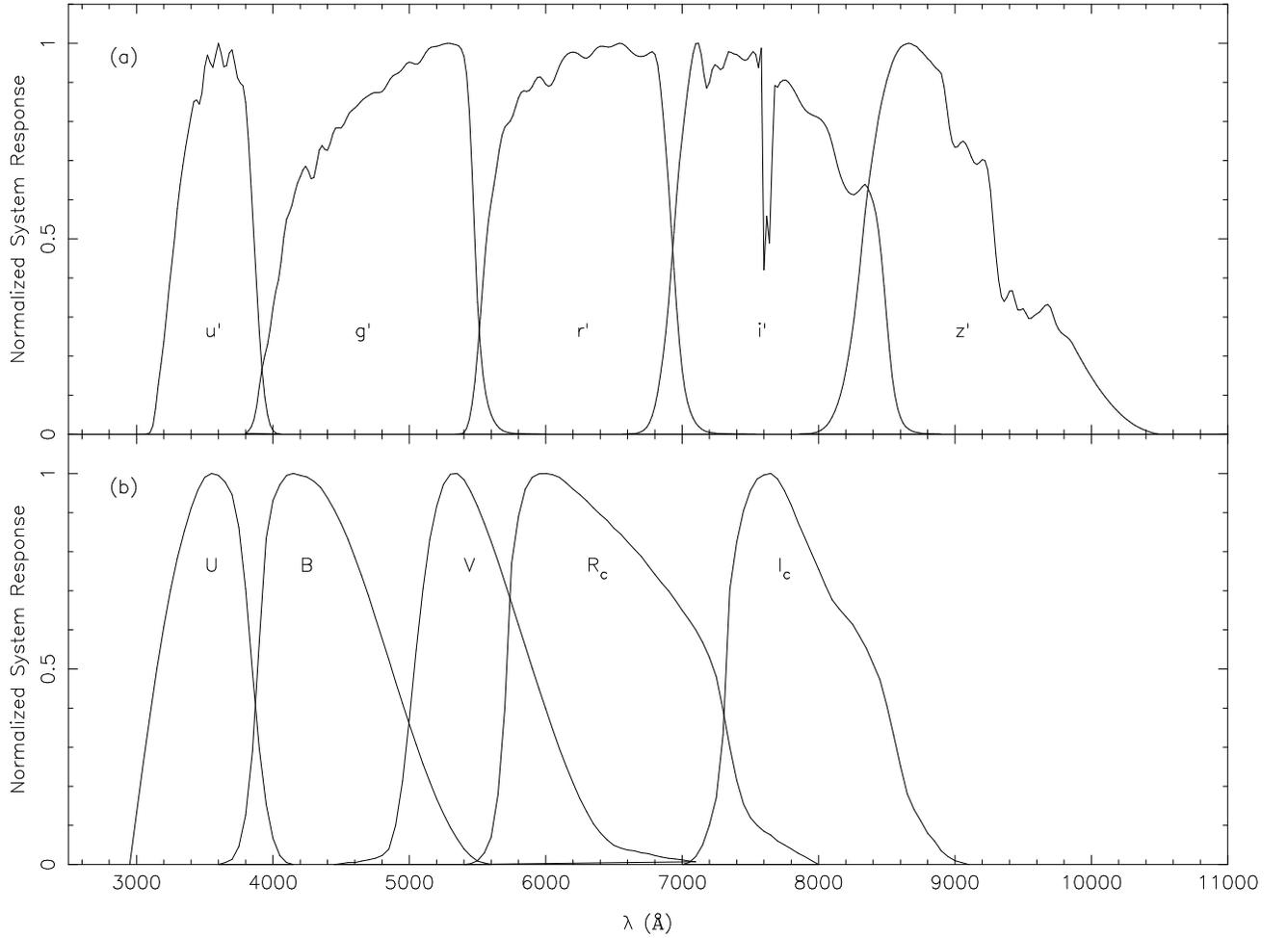}
\caption{The (normalized) responses of the $u'g'r'i'z'$ system bandpasses 
(at 1.2 airmasses of extinction) compared those of the the 
Johnson-Morgan-Cousins ($UBVR_{\rm c}I_{\rm c}$) system. (Filter curves for 
the Johnson-Morgan $UBV$ filters and for the Cousins $R_{\rm c}I_{\rm c}$ 
filters were obtained from The General Catalogue of Photometric Data at 
{\tt http://obswww.unige.ch/gcpd/gcpd.html};
\cite{MMH97}.)  
\label{ugrizUBVRIfilters}}

\end{figure}

The five filters of the $u'g'r'i'z'$ system have effective wavelengths
of 3540 \AA, 4750 \AA, 6222 \AA, 7632 \AA, and 9049 \AA, respectively,
at 1.2 airmasses.\footnote{Note that the $g'$ filter has been
determined to have an effective wavelength 20 \AA\, bluer than that
originally quoted by \citet{fuk96}.}  They cover the entire wavelength
range of the combined atmosphere+CCD response and their construction is
described by \citet{fuk96}.  Also shown in that paper (their Figure~1)
are the designed response curves for the filters multiplied by the QE curve 
of a thinned, UV-AR coated Tektronix TK1024 CCD, similar to the detector
that was used in the development of this standard system.  The
$u'g'r'i'z'$ filters have sharp cutoffs by design.  The passbands were
selected to exclude the strongest night-sky lines; for example
\ion{O}{1} ($\lambda$5577) and
\ion{Hg}{1} ($\lambda$5461).  The bulk of the $u'$ band response is
blueward of the Balmer discontinuity which, when combined with the
$g'$ filter, yields high sensitivity to the magnitude of the Balmer
jump but at a cost of lower throughput for the narrower $u'$ filter
(compared with Johnson $U$).  In Figure~\ref{usnoResponse}, we show
the filter responses multiplied by the sensitivity data for a CCD
similar to that used to set up the $u'g'r'i'z'$ system.  These curves
represent the expected total quantum efficiencies of the filter
transmissions, the QE of the CCD surface, and the reflections from two
aluminum mirror surfaces.  The two sets of response curves shown are
for the case without atmospheric extinction (upper curves) and as
modified for typical extinctions at 1.2 airmasses (lower curves).
Figure~\ref{ugrizUBVRIfilters} compares the (normalized) $UBVR_{\rm
c}I_{\rm c}$ filter curves with those of the $u'g'r'i'z'$ system.

The filter transmission data as measured by the Japan Participation 
Group within the SDSS, the filter manufacturing specifications and the 
CCD+filter response curves are available on-line at
{\tt http://home.fnal.gov/$\sim$dtucker/ugriz/index.html}.  These
curves, as well as the other $u'g'r'i'z'$ links from this page will
be updated as needed.

\placefigure{usnoResponse}

\placefigure{ugrizUBVRIfilters}

This project used 183 nights of telescope time spanning a 22 month (24
lunation) period beginning in March 1998.  The raw statistics for the
success of each observing run are in Table~\ref{obsruns} where the
first two columns give the year and month of the observing session
followed by the UT date.  The third and fourth columns give the number
of nights allocated on the telescope and the number of those nights
that were clear, where ``clear'' is defined as no clouds were seen in
the sky by the observer for a stretch greater than three hours at a
time.  As a consequence, we collected data on some nights indicated as
``clear'' that later proved not to be usable. The last column gives
the total number of usable $u'g'r'i'z'$ observations from each
observing run where one observation indicates one star observed once
in each of the five filters.

\placetable{obsruns}

\section{The Stars \& Observing Strategy}

In order to have the standard star system in place for the start of
SDSS science operations, as required for follow-up spectroscopy
target selection algorithms, it was necessary to save time during 
establishment of the network of standard stars.  This was done by 
making use of previous work on
standard stars so that variable stars in the initial fields were
already identified.  A preliminary list of 63 standard
candidates was derived from the work of \citet{TG76},
\citet{OG83}, and \citet{oke90}.  This list was supplemented using
the work of \citet{sand64}, \citet{vee74}, \citet{stone77}, and
\citet{kent85}, and then pared to 36 stars using a magnitude cut.  The
remaining stars were then heavily supplemented using equatorial
$UBVR_{\rm c}I_{\rm c}$ standard stars (\citealt{lan73},
\citeyear{lan83}, \citeyear{lan92}) which served to fill in gaps in
right ascension and provide potential color pairs for secondary
extinction terms.  Additional red stars were obtained from the USNO
photometry program (H. Harris, private communication, 1998).  At the
beginning, our preliminary list contained roughly 200 candidate stars.
In the end, the list was trimmed to just those stars with ten or more
observations (in each of the five filters) in the current program, or four 
or more observations in our program provided
they had been used as standards in other systems with at least 10
observations to indicate they were not variables.  In the end, 164
stars were observed often enough to be included in the reduction
process for the final network.  Of these, 158 were retained
for the final catalog of standards defining the $u'g'r'i'z'$
system.

The primary goals for the SDSS are large scale structure studies using
galaxies and QSOs so this first version of the standard network is
limited, for the most part, to stars bluer than about M0 to avoid the
strengthening metal bands and flare stars. Further, most of the survey
area is around the North Celestial Pole with a limited area near the
celestial equator, so a heavy emphasis was placed on stars in the
northern hemisphere.  Expansion of the network to the southern
hemisphere and to redder stars is underway (see \S6).

Rather than selecting some fiducial spectral type to have null colors
(as in the $UBVR_{\rm c}I_{\rm c}$ system), our aim was to set the
$u'g'r'i'z'$ standard star network on the $AB$ system.  Recall that, in
the $AB$ system, a monochromatic magnitude is defined such that
\begin{equation}
AB_{\nu} = -2.5~\log f_{\nu} - 48.60 ,
\end{equation}
where $f_{\nu}$ is the flux per unit frequency from an object in
ergs~s$^{-1}$~cm$^{-2}$~Hz$^{-1}$; thus,
\begin{equation}
f_{\nu}({\rm Jy}) = 3631~{\rm dex}(-0.4AB_{\nu}) 
\end{equation}
\citep{OG83,fuk96}.

As noted, the $AB$ is, strictly speaking, a {\em monochromatic\/}
system, defining a magnitude for a single frequency $\nu$.  The
$u'g'r'i'z'$ system, however, is very much a {\em broadband\/} filter
system.  Tying a broadband system to a monochromatic flux is
complicated by the existence of stellar absorption lines and by the
fact that the mean wavelength of a broadband filter depends on a
given star's color.  As a compromise solution, we follow \citet{fuk96}'s
lead, and define an $AB$ {\em broadband} magnitude by the following
equation:
\begin{equation}
m = -2.5\log\frac{ \int d(\log \nu) f_{\nu} S_{\nu} }{ \int d(\log \nu) S_{\nu} } - 48.60 , \label{eq:broadbandAB}
\end{equation}
where $f_{\nu}$ is the energy flux per unit frequency on the atmosphere
and $S_{\nu}$ is the system response.

To zeropoint the $u'g'r'i'z'$ system, we used the synthetic $AB$
$u'g'r'i'z'$ magnitudes of the F subdwarf BD+17$\arcdeg$4708 as
calculated by \citet{fuk96}.  Using equation~\ref{eq:broadbandAB},
\citet{fuk96} computed this star's $AB$ broadband magnitudes by convolving the
$u'g'r'i'z'$ system response (Fig.~\ref{usnoResponse}) with the
Vega-calibrated spectrophotometry of BD+17$\arcdeg$4708.  Our
observations of BD+17$\arcdeg$4708 were then forced to match these
synthetic magnitudes and all of the standard stars were scaled to this
zeropoint.  Thus there is a formally defined relationship between
$u'g'r'i'z'$ magnitudes and photon flux.

Initial estimates (Eisenstein, private communication; Finkbeiner,
private communication) indicate that the present network deviates from
a true $AB$ broadband system by no more than about 10\% in $u'$ and
$z'$ and 5\% in $g'r'i'$.  These systematic errors are due to
uncertainties in the absolute calibration of the synthetic
$u'g'r'i'z'$ magnitudes of BD+17$\arcdeg$4708, and include
uncertainties:
\begin{itemize}
\item in the USNO filter transmissions, 
\item in the CCD response, 
\item in the atmospheric correction to the filter curves,  
\item in the relative calibration of BD+17$\arcdeg$4708
	to Vega, and 
\item in the absolute calibration of Vega
\end{itemize}
Since the deviations from a true $AB$ system are due to
zeropoint magnitude offsets in the absolute calibration of
BD+17$\arcdeg$4708 and not due to linear color shifts within the
network of stars itself, future corrections towards a $AB$ system ---
if warranted --- should only entail a small additive constant in the
standard star magnitudes in each filter band, once these constants
have been well determined.

In practice, during the early stages of this program, we needed more
than one star to cover the sky for those times when BD+17$\arcdeg$4708
was not visible, so two other F subdwarfs were chosen to be reference
stars to supplement BD+17$\arcdeg$4708: BD+26$\arcdeg$2606 and
BD+21$\arcdeg$0607 (Table~\ref{fundamentals}).  BD+26$\arcdeg$2606 has
excellent spectrophotometry relative to BD+17$\arcdeg$4708, and
BD+21$\arcdeg$0607 has excellent photometric measurements in the
Thuan-Gunn system. Thus, all three of these stars were used as 
fundamental calibration stars during the setup of the $u'g'r'i'z'$
standard star network.  Note, however, that the final system zeropoint is
tied solely to the synthetic photometry of BD+17$\arcdeg$4708.

\placetable{fundamentals}

\begin{figure}
\figurenum{4}
\epsscale{0.75}
\includegraphics[angle=-90,scale=.75]{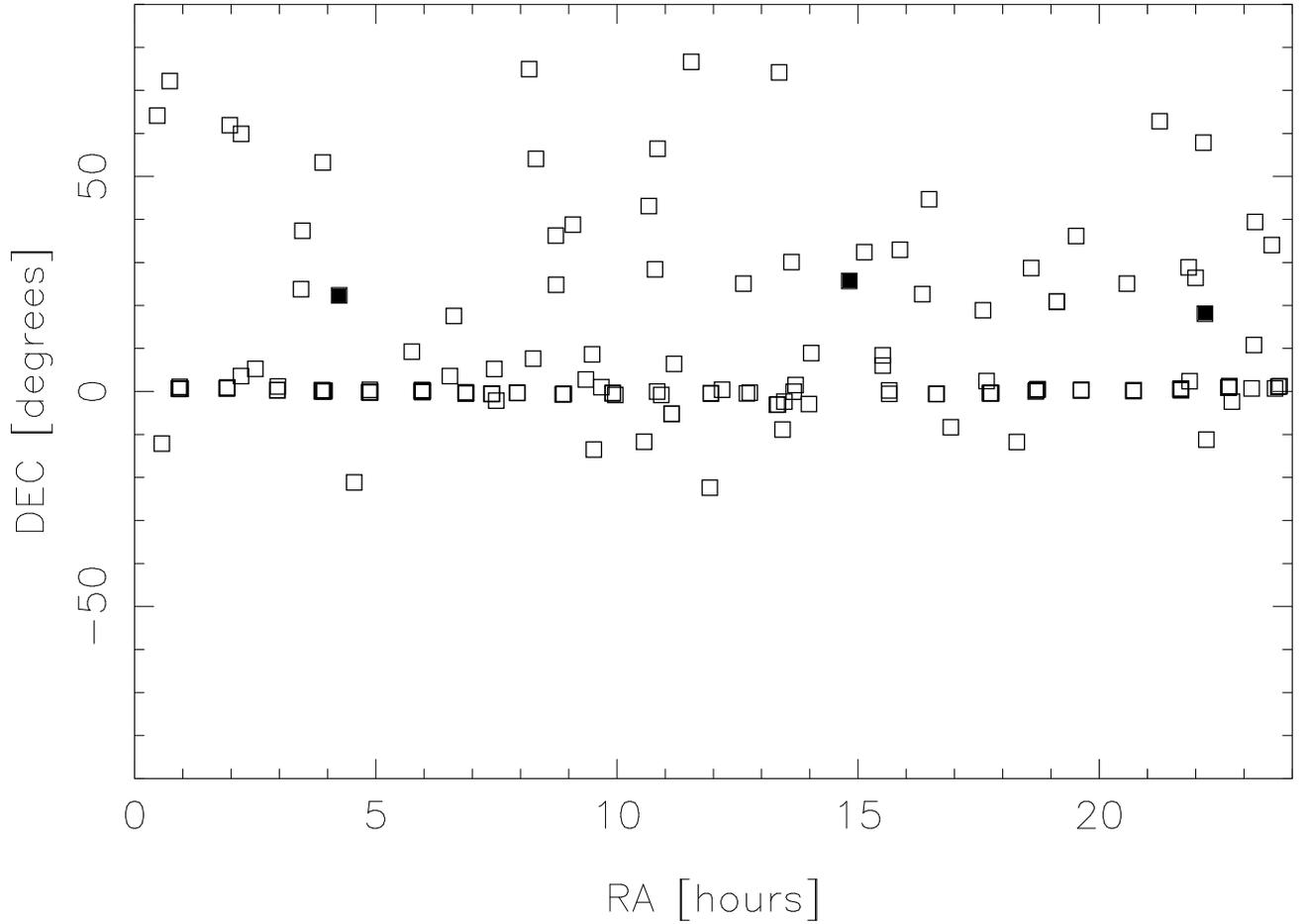}
\caption{The distribution of the primary standards in right ascension and 
declination.  Clearly seen are the clustering of stars near the celestial 
equator and the relative dearth of standards in the southern hemisphere.  
Most of the equatorial fields contain multiple stars therefore, though there 
are 158 stars in the system, there are not as many individual points on this 
plot.  The three fundamental standards --- BD+17$\arcdeg$4708, 
BD+26$\arcdeg$2606, and BD+21$\arcdeg$0607 are indicated by the filled 
symbols. 
\label{RAvsDEC}}
\end{figure}

At the telescope, we tried to observe one of the three
fundamental stars at least every 90 minutes to determine the zero
points and judge stability for each night.
In addition to the fundamental standards, the target list of
primary standards was chosen to maximize the color and airmass ranges
for each night.  In general, two or three primary fields were
observed several times to monitor extinction manually at the telescope
over the course of a night.  These values were compared with the
``all--sky'' extinction values determined later by the reduction
software using all observations for the night.  Additional fields were
also observed in a random order to provide a good color spread near
the meridian and at high airmass.  We attempted to observe each of
these fields two or three times a night with at least one hour between
repeated observations.  On the next usable night of the same observing
run, a different set of manual extinction fields was selected and
different primary candidate fields were chosen with some, but not
complete, overlap.  This method reduces dependence on any particular
set of extinction stars.  The equatorial fields were generally used
for extinction to maximize the airmass range and leverage the effect
of the color terms.  The sky distribution of the final set of stars is
shown in Figure~\ref{RAvsDEC}.

\placefigure{RAvsDEC}

Finally, typically several additional stars within each candidate field
were tagged as ``extra'' or ``monitor'' stars.  Although these stars
were excluded from the $u'g'r'i'z'$ standard star network,
they are being calibrated in the $u'g'r'i'z'$ system and, after
culling variables, will form the basis of a future catalog of
supplementary $u'g'r'i'z'$ standards useful for photometric
calibration of data from large-format CCD mosaic imaging cameras
\citep{smith02}.

\section{Reductions}

\subsection{Software}

The reductions for this standard system were performed using the
``Monitor Telescope Pipeline'' ({\tt MTPIPE}), a suite of code written
in the SDSS software environment \citep{sto95,ser96}.  For the development 
of the standard star network, the pipeline reductions for each night occurred 
in three steps --- {\tt preMtFrames}, {\tt mtFrames}, and {\tt excal} --- 
briefly described here.  A more detailed description can be found in an 
upcoming paper \citep{dlt00}.

The first package, {\tt preMtFrames}, creates the directory structure
for the reduction of a night's data, including parameter files needed
as input for the other three packages, and runs quality assurance
tests on the raw data.  It identifies the image type (e.g., bias
frame, twilight flat, standard field), matches the frame name to a
list of approved standard star field names, verifies that a full set
of frames ($u'g'r'i'z'$) are present for each field, and creates
quality assurance histograms for the bias and flat field frames.

The next package, {\tt mtFrames}, is the image processing
portion of the software.  This package creates median filtered bias
and flat frames, applies them to the images, and then extracts
aperture photometry on all objects found in each image.  
The candidate standard stars are measured in a 24-arcsec diameter aperture.  
This large size was selected to avoid problems associated with defocussing 
the brightest stars, required for some of the observations.

The third package, {\tt excal}, takes the output of {\tt mtFrames} and
identifies the individual primary candidate and fundamental stars within each
field, then calculates the photometric zeropoint and the atmospheric
extinction using the instrumental magnitudes for these stars as input.
Before the standard star network was calibrated, we used the candidate 
primary stars as extinction standards and also solved for
their best-fit magnitudes based upon that night's observations.  In
this mode of operation, only the three fundamental stars ---
BD+17$\arcdeg$4708, BD+26$\arcdeg$2606 and BD+21$\arcdeg$0607 --- had
fixed magnitudes; observations of one or more of these three set the
photometric zeropoint for the night.

Two additional {\tt mtpipe} packages --- {\tt solve\_network} and {\tt
superExcal} --- were used for the final calibration of the primary
standard star network.  These are similar to {\tt excal} in that they
solve for the photometric parameters of a set of data and for the
best-fit magnitudes of the primary standards.  They differ from {\tt
excal} in that they use a single star --- BD+17$\arcdeg$4708 --- to
set the zeropoint for the photometric solution.  BD+17$\arcdeg$4708 is
{\em defined\/} to have the magnitudes given in
Table~\ref{fundamentals} and sets the zeropoint for the SDSS standard
star network.  Although both {\tt solve\_network} and {\tt superExcal}
perform a similar task, they differ from each other in that {\tt
superExcal} is largely an outgrowth of the code in {\tt excal} and its
least squares solver, but generalized to run on multiple nights of
data; {\tt solve\_network}, however, was written completely
independently of the excal code by one of us (MWR).  As such, {\tt
solve\_network} has provided a useful independent check of our {\tt
superExcal} results.

To calibrate the standard stars, we took the following steps:
\begin{enumerate}
\item We ran {\tt preMtFrames, mtFrames}, and {\tt excal} on each night of
      data from the USNO 1.0-m telescope.
\item From the output of {\tt excal} and from observer notes at the telescope,
      we determined which nights were photometric.
\item Using the output of {\tt mtFrames} and {\tt excal}, we ran 
      {\tt superExcal} for the final version of the calibrations.
\item Using the output of {\tt mtFrames} and {\tt excal}, we ran the 
      {\tt solve\_network} code as an independent check of our 
      {\tt superExcal} results.  
\item Finally, we applied a small (0.00--0.04mag) red leak correction
      to the $u'-g'$ colors, to remove the effects of the $u'$ filter
      red leaks (see \S4.3).
\end{enumerate}

\subsection{The Photometric Equations}

The equations used to recover the $u'g'r'i'z'$ system have the form:  
\begin{eqnarray}
u'_{\rm inst} & = & u'_{\rm o} + a_{u} + b_{u} (u'-g')_{\rm o} + k_{u} X 
     + c_{u} [(u'-g')_{\rm o} - (u'-g')_{\rm o,zp}] [X-X_{\rm zp}]\;, \\
g'_{\rm inst} & = & g'_{\rm o} + a_{g} + b_{g} (g'-r')_{\rm o} + k_{g} X 
     + c_{g} [(g'-r')_{\rm o} - (g'-r')_{\rm o,zp}] [X-X_{\rm zp}]\;, \\
r'_{\rm inst} & = & r'_{\rm o} + a_{r} + b_{r} (r'-i')_{\rm o} + k_{r} X 
     + c_{r} [(r'-i')_{\rm o} - (r'-i')_{\rm o,zp}] [X-X_{\rm zp}]\;, \\
i'_{\rm inst} & = & i'_{\rm o} + a_{i} + b_{i} (i'-z')_{\rm o} + k_{i} X 
     + c_{i} [(i'-z')_{\rm o} - (i'-z')_{\rm o,zp}] [X-X_{\rm zp}]\;, \\
z'_{\rm inst} & = & z'_{\rm o} + a_{z} + b_{z} (i'-z')_{\rm o} + k_{z} X 
     + c_{z} [(i'-z')_{\rm o} - (i'-z')_{\rm o,zp}] [X-X_{\rm zp}]\;.
\end{eqnarray}
Taking the $g'$ equation as an example, we note that $g'_{\rm inst}$
is the measured instrumental magnitude, $g'_{\rm o}$ is the
extra-atmospheric magnitude, $(g'-r')_{\rm o}$ is the
extra-atmospheric color, $a_{g}$ is the nightly zero point, $k_{g}$ is
the first order extinction coefficient, $b_{g}$ is the system transform
coefficient, and $c_{g}$ is the second order (color) extinction
coefficient.  The airmass of observation, $X$, is as defined by
\citet{bem04}.

The zeropoint constants, $X_{\rm zp}$ and $(g'-r')_{\rm o,zp}$, used
in the second order extinction term, were defined, respectively, to be the average
standard star observation airmass $<X>$ = 1.3 and the ``cosmic
color,'' as listed in Table~\ref{cosmic_colors}.  The
cosmic color values were derived from 4428 objects with $19 < r' < 20$
in SDSS survey run 752, camera column 3, fields 11-100 [see
\citet{GCRS98} for a description of the SDSS survey camera].  This
area is on the celestial equator at a Galactic latitude of about
42$\arcdeg$.  The use of these zeropoint constants permits setting
$c_{g}$ to zero without affecting the values of the other terms in the
photometric equation ($a_{g}$, $b_{g}$, and $k_{g}$), thus simplifying
the photometric equations for projects not requiring the highest
photometric accuracy.

\placetable{cosmic_colors}

The {\tt excal} and {\tt superExcal} packages solve each of the
photometric equations (eqs.~4--8) iteratively, performing the
following loop:
\begin{enumerate}

\item Feed into the equations the current estimates for the 
extra-atmospheric magnitudes, colors, and photometric coefficients
(e.g., for eq.~5, $g'_{\rm o}$, $r'_{\rm o}$, $a_{g}$, $b_{g}$,
$c_{g}$, and $k_{g}$).  For the first iteration of this loop, initial
estimates for these parameters are fed into the equations.

\item Solve each of the equations in turn for the extra-atmospheric
magnitude and the non-color-dependent coefficients (e.g., $g'_{\rm
o}$, $a_{g}$, and $k_{g}$), keeping the color-dependent coefficients
(e.g., $b_{g}$, $c_{g}$) fixed.

\item Solve each of the equations in turn for the two color-dependent 
coefficients (e.g., $b_{g}$, $c_{g}$), keeping the extra-atmospheric
magnitude and the non-color-dependent coefficients fixed (e.g.,
$g'_{\rm o}$, $a_{g}$, and $k_{g}$).

\item Permit the user to delete (or undelete) outliers interactively.

\end{enumerate}

This loop will be performed as long as the user continues to delete
(or undelete) observations interactively from the solution.  (To ensure
that the user has finished modifying the dataset, this loop will run
an additional three times after the last user modification before
outputting the final photometric solutions.)

One can also choose to fix any of the photometric coefficients to a
pre-set value and {\em not\/} solve for it.  For instance, in setting
up the standard star network, we chose to fix the system transform
coefficients (e.g., $b_{g}$ in eq.~5) to zero, since the USNO~1.0-m,
its CCD, and its filters are the defining instruments of the
$u'g'r'i'z'$ system.

Finally, we note that equations~4--8, which are fit by {\tt excal}
and {\tt superExcal}, take the standard magnitudes and colors and
convert them to instrumental magnitudes.  
The inverse equations are:
\begin{eqnarray}
u'_{\rm o} & = &  u'_{\rm inst} + \hat{a}_{u} + \hat{b}_{u} (u'-g')_{\rm inst} 
                                + \hat{k}_{u} X \nonumber \\ 
       & + & \hat{c}_{u} [(u'-g')_{\rm inst} - (u'-g')_{\rm inst,zp}] [X-X_{\rm zp}]\;, \\
g'_{\rm o} & = &  g'_{\rm inst} + \hat{a}_{g} + \hat{b}_{g} (g'-r')_{\rm inst} 
                                + \hat{k}_{g} X \nonumber \\ 
       & + & \hat{c}_{g} [(g'-r')_{\rm inst} - (g'-r')_{\rm inst,zp}] [X-X_{\rm zp}]\;, \\
r'_{\rm o} & = &  r'_{\rm inst} + \hat{a}_{r} + \hat{b}_{r} (r'-i')_{\rm inst} 
                                + \hat{k}_{r} X \nonumber \\ 
       & + & \hat{c}_{r} [(r'-i')_{\rm inst} - (r'-i')_{\rm inst,zp}] [X-X_{\rm zp}]\;, \\
i'_{\rm o} & = &  i'_{\rm inst} + \hat{a}_{i} + \hat{b}_{i} (i'-z')_{\rm inst} 
                                + \hat{k}_{i} X \nonumber \\ 
       & + & \hat{c}_{i} [(i'-z')_{\rm inst} - (i'-z')_{\rm inst,zp}] [X-X_{\rm zp}]\;, \\
z'_{\rm o} & = &  z'_{\rm inst} + \hat{a}_{z} + \hat{b}_{z} (i'-z')_{\rm inst} 
                                + \hat{k}_{z} X \nonumber \\ 
       & + & \hat{c}_{z} [(i'-z')_{\rm inst} - (i'-z')_{\rm inst,zp}] [X-X_{\rm zp}]\;.
\end{eqnarray}
Note that the inverse coefficients have carets, indicating that
they do not necessarily have the same values as the 
direct coefficients of equations~4--8. For a filter $x$ with filter
$y$ as its color index conjugate --- i.e., filter $x$'s color index is
$(x - y)$ --- the conversions from {\em direct\/} to {\em inverse\/}
coefficients are
\begin{eqnarray}
\hat{a}_x & = & a_x - b_x (a_x - a_y)\;, \\
\hat{b}_x & = & b_x\;, \\
\hat{c}_x & = & c_x\;, \\
\hat{k}_x & = & k_x - b_x (k_x - k_y)\;.
\end{eqnarray}
Again, using the SDSS $g'$ filter as a concrete example,
\begin{eqnarray}
\hat{a}_g & = & a_g - b_g (a_g - a_r)\;,\\
\hat{b}_g & = & b_g\;, \\
\hat{c}_g & = & c_g\;, \\
\hat{k}_g & = & k_g - b_g (k_g - k_r)\;.
\end{eqnarray}
Furthermore, we note that for the {\em inverse\/} equations, we must use
an instrumental form of the color zeropoint in the second order extinction
term.  Again, for a generic filter $x$ with color index $(x - y)$, 
\begin{equation}
(x'-y')_{\rm inst,zp} = (x'-y')_{\rm o,zp} + (a_x - a_y) + (k_x - k_y) X\;,
\end{equation}
and, for our concrete example, 
\begin{equation}
(g'-r')_{\rm inst,zp} = (g'-r')_{\rm o,zp} + (a_g - a_r) + (k_g - k_r) X\;.
\end{equation}

\subsection{The Red Leak Correction}

There are two small red leaks associated with the $u'$ filter ---  
one at 8120 \AA\, and another, larger leak beyond 10,000 \AA.  The latter
red leak is largely suppressed by the USNO 1.0-m CCD's low quantum efficiency
at these near-infrared wavelengths.

If we define a flux sensitivity quantity, $Q$, by
\begin{equation}
Q \equiv \int d(\ln \nu) S_{\nu}, 
\end{equation}
where $\nu$ is the frequency and $S_{\nu}$ is the system quantum
efficiency \citep[ eq.~5]{fuk96}, we can calculate the effects of the
two red leaks relative to a non-red-leak $u'$ passband.  Measured for
an airmass of 1.2, we find for the non-red-leak $u'$ passband, the
8120 \AA\, red leak, and the $>$10,000 \AA\, red leak, respectively:
\begin{eqnarray}
Q_{u'} & = & 1.842 \times 10^{-2} \nonumber \\
Q_1    & = & 1.009 \times 10^{-6} \\
Q_2    & = & 0.783 \times 10^{-7} \nonumber
\end{eqnarray}
The effect of these red leaks in magnitudes can then be calculated
by
\begin{equation}
\Delta_i = -2.5 \log ( Q_i/Q_{u'} ) ,
\end{equation}
where $i=$ 1 or 2.  We find:
\begin{eqnarray}
\Delta_1 & = & 10.654 \nonumber \\
\Delta_2 & = & 10.929
\end{eqnarray}  
Combining these results, we can calculate the correction we need to
apply to an arbitrary $u'-x'$ color of a star (where $x'$ is an 
arbitrary filter) to remove the effects of these two
red leaks:
\begin{eqnarray}
(u'-x')_{\rm true} & = & (u'-x')_{\rm obs} \nonumber \\
                   & + & 2.5 \log  \left(
                     1 + 0.4~{\rm dex}[(u'-i')_{\rm true} - \Delta_1]
                       + 0.4~{\rm dex}[(u'-z')_{\rm true} - \Delta_2]
                     \right) , \label{redleakcorr1}
\end{eqnarray}
where $(u'-x')_{\rm obs}$ is $(u'-x')$ before the red leak correction
and $(u'-x')_{\rm true}$ is $(u'-x')$ after the red leak correction.
Note that $(u'-i')_{\rm true}$ and $(u'-z')_{\rm true}$ are merely
$(u'-x')_{\rm true}$ for $x' = i'$ and $x' = z'$, respectively.

Note that equation~\ref{redleakcorr1} is an implicit equation and must
be solved iteratively.  Fortunately, for cases like ours in which the red leak
correction is small, we can use the following approximation:
\begin{eqnarray}
(u'-x')_{\rm true} & = & (u'-x')_{\rm obs} \nonumber \\
                   & + & 1.086  \times \left(
                     {\rm dex}[ 0.4 ((u'-i')_{\rm obs} - \Delta_1)]
                   + {\rm dex}[ 0.4 ((u'-z')_{\rm obs} - \Delta_2)]
                     \right) , \label{redleakcorr2}
\end{eqnarray}
where we were able to replace $(u'-i')_{\rm true}$ with $(u'-i')_{\rm
obs}$ and $(u'-z')_{\rm true}$ with $(u'-z')_{\rm obs}$ due to the
smallness of the red leak corrections.  In \S~5, where we present the
magnitudes and colors for the $u'g'r'i'z'$ primary standards, we use
equation~\ref{redleakcorr2} to correct the $u'-g'$ colors.

\subsection{Data}

Both {\tt excal} and {\tt superExcal} solve for first order extinction
coefficients that vary over the course of the night; the extinction is
determined for individual segments of time called ``solution
time-blocks.''  The first order extinction coefficients for each night
used in the {\tt superExcal} photometric solution are given in
Table~\ref{kterms}.  These are arranged by UT date and block
(YYMMDD-B) with the corresponding Modified Julian Date for the
midpoint of each solution time-block (mid-mjd), rounded to the nearest
0.01 day.\footnote{The Modified Julian Date is defined by the relation
MJD $\equiv$ JD $-$ 2400000.5, where JD is the Julian Date.}  These
are listed in the first two columns of the table.  The extinction
coefficients and associated uncertainties are given in columns 3--7.
These values were determined as part of the {\tt superExcal} solution.
The default solution block within {\tt superExcal} was set at 3 hours.
Each night was examined for obvious breaks --- east vs.\ west; the
moon; breaks around possible clouds; before and after midnight --- and
these were given priority for dividing the night into reduction
blocks.  In the absence of obvious breaks, the default solution block
within {\tt MTPIPE}, 3 hours, was used.  Further, each block had to
contain a minimum of 10 observations per passband (about 1.5 hours of
observing time), after all rejections, to be kept in the final
solution.  In the end, 109 usable blocks on 61 nights were determined
to be photometric and contained the minimum number of observations to
be included in the system solution.

\placetable{kterms}

\begin{figure}
\figurenum{5}
\epsscale{0.75}
\plotone{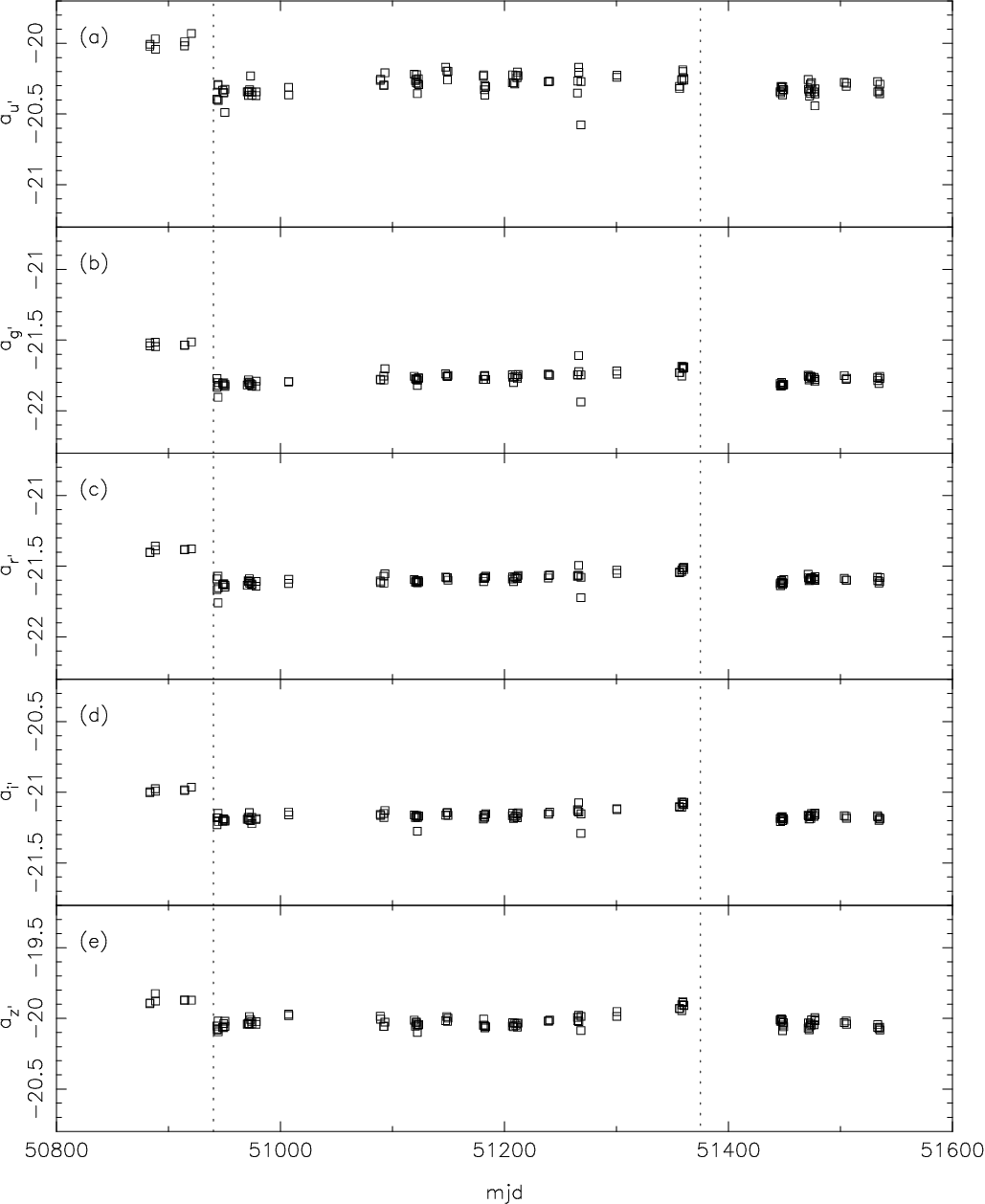}
\caption{The photometric zeropoints calculated for each night of data used 
in developing the standard star network.  The mirrors were re-aluminized 
twice during our program --- once after the second month of observations and 
again prior to the last four months of observations (denoted by the dotted 
vertical lines).  The first re-aluminization is clearly visible as a break 
in the zero point values while the second break is less obvious.  The two 
large gaps ($\approx$mjd5100--51100 and mjd51350--51450) correspond to the 
two monsoon seasons (summers) in northern Arizona.
\label{a_vs_mjd}}
\end{figure}

The zeropoints for each of the 61 usable nights are shown in
Figure~\ref{a_vs_mjd}.  These are calculated by {\tt superExcal}
once for each block of time, from fundamental standard observations.
As shown, most nights have similar zero points.  Clearly seen in the
figure are the times when the mirrors were re-aluminized: once after
the second month and again prior to the last four months of
observations.  A few outlying points correspond with high extinction
values but the derived magnitudes for these nights are within the
2$\sigma$ error limits for stars observed on those nights.

\placefigure{a_vs_mjd}

\begin{figure}
\figurenum{6}
\epsscale{0.75}
\plotone{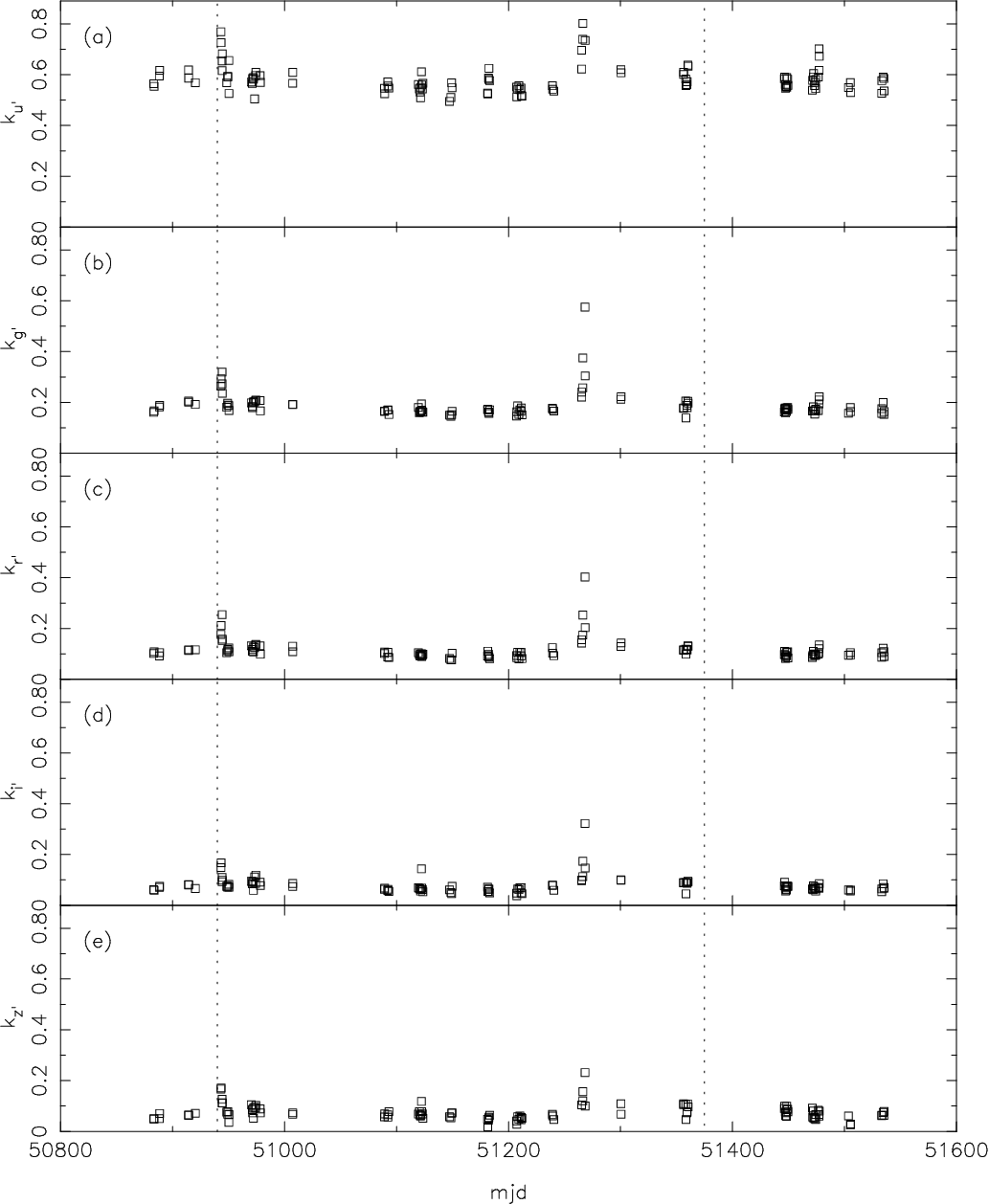}
\caption{The primary extinction coefficients for each 
block of data used in developing the standard star network.  The five
nights with high extinction values (mjd50943, 51265, 51266, 51268 and,
51477) correspond to the night with low zeropoints. The times when the
mirror was re-aluminized are denoted by the two dotted vertical lines.
\label{k_vs_mjd}}
\end{figure}

The primary extinction coefficients (Table~\ref{kterms}) for each 
block are plotted in Figure~\ref{k_vs_mjd}.  As seen, the extinction values 
were generally stable during the program.  Three nights with higher 
than normal extinction values (and low zero points) are seen.  However, as 
mentioned above, the final magnitude values for each of the stars observed 
during these times were within the accepted limits.

\placefigure{k_vs_mjd}

Secondary extinction (color term) values are given in Table~\ref{cterms}.  
Calculating these values is an option in the pipeline code.  For the 
system setup, we assumed there would be one fixed value for all nights 
and solved for these coefficients.  This correction is minor compared 
with the primary extinction terms so our assumption of a single 
value does not impact the solution.  (For comparison, we also list
the values for the secondary extinction coefficients obtained by
{\tt solve\_network}.)

\placetable{cterms}

The residuals of each observation in the network solution were examined
for trends, as a function of several different variables, within the 
final network solution.  The plots of these test results are available on 
our $u'g'r'i'z'$ web site 
{\tt http://home.fnal.gov/$\sim$dtucker/ugriz/index.html}.  To
summarize, we found no apparent trends in the residuals as a function of:

\begin{itemize}
\item time (over the course of the program),
\item airmass,
\item magnitude,
\item color,
\item product of color times airmass,
\item right ascension,
\item declination,
\item hour angle,
\item ambient temperature of observation.
\end{itemize}

We did see an increase in the scatter of the residuals in the $u'$ filter
at fainter magnitudes and for redder stars.
This was caused by the tailored exposure times used on each field with
multiple candidate stars.  Since the exposure times for each standard field 
were tailored for the brightest candidate, the second, third, fourth ... star 
in each field will be increasingly under-exposed resulting in the scatter.  
We also noticed an increase in scatter for the bluer star residuals in the
$z'$ filter; also an artifact of the tailored exposure lengths. 

\begin{figure}
\figurenum{7}
\epsscale{0.75}
\plotone{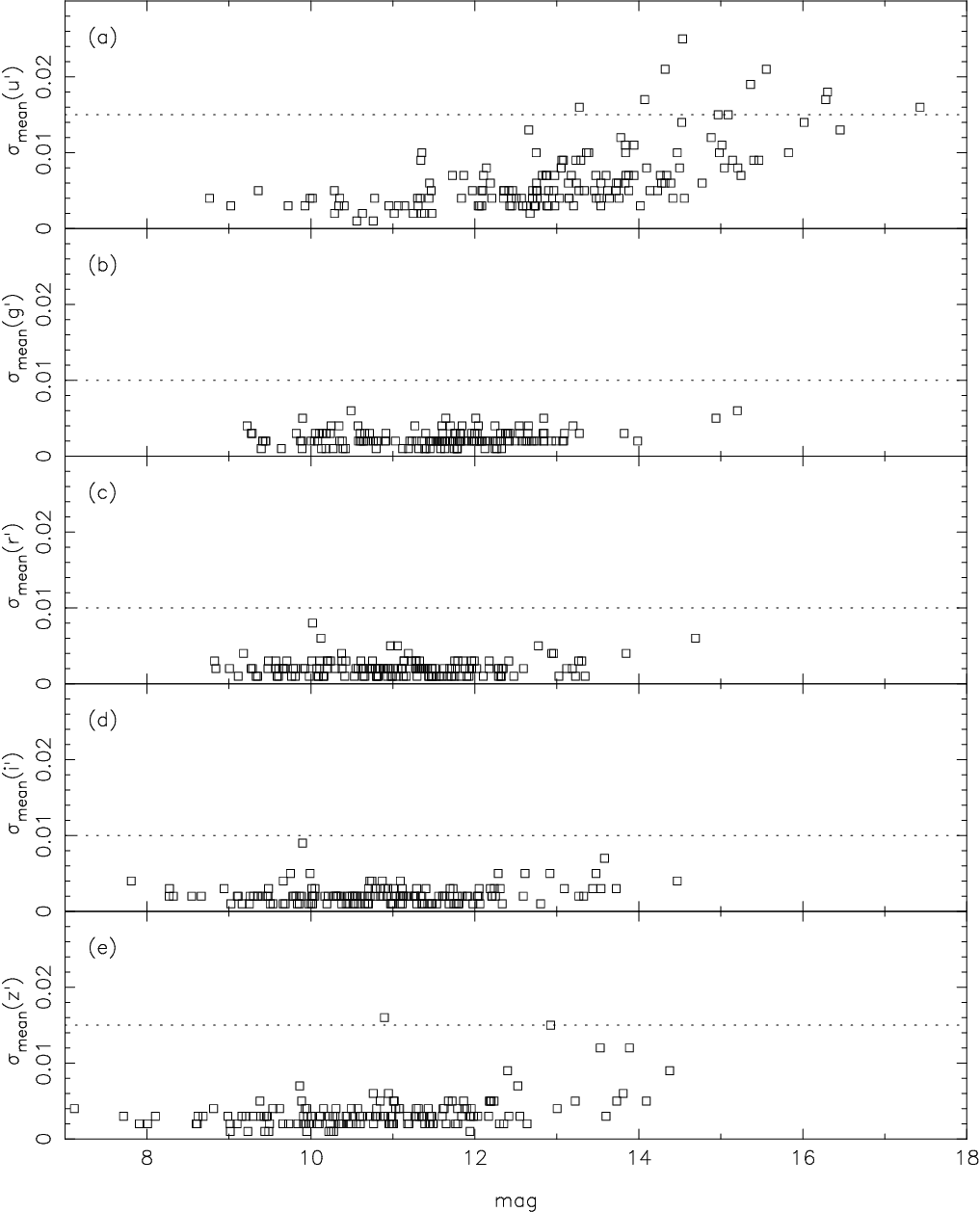}
\caption{Mean error vs.\ magnitude
($\sigma_{\rm mean} (u')$ vs.\ $u'$, $\sigma_{\rm mean} (g')$ vs.\ $g'$,
$\sigma_{\rm mean} (r')$ vs.\ $r'$, $\sigma_{\rm mean} (i')$ vs.\ $i'$,
$\sigma_{\rm mean} (z')$ vs.\ $z'$).  The horizontal dashed lines are the
survey requirement for the standard network to meet.
\label{meanerror_vs_mag}}
\end{figure}

\begin{figure}
\figurenum{8}
\epsscale{0.75}
\plotone{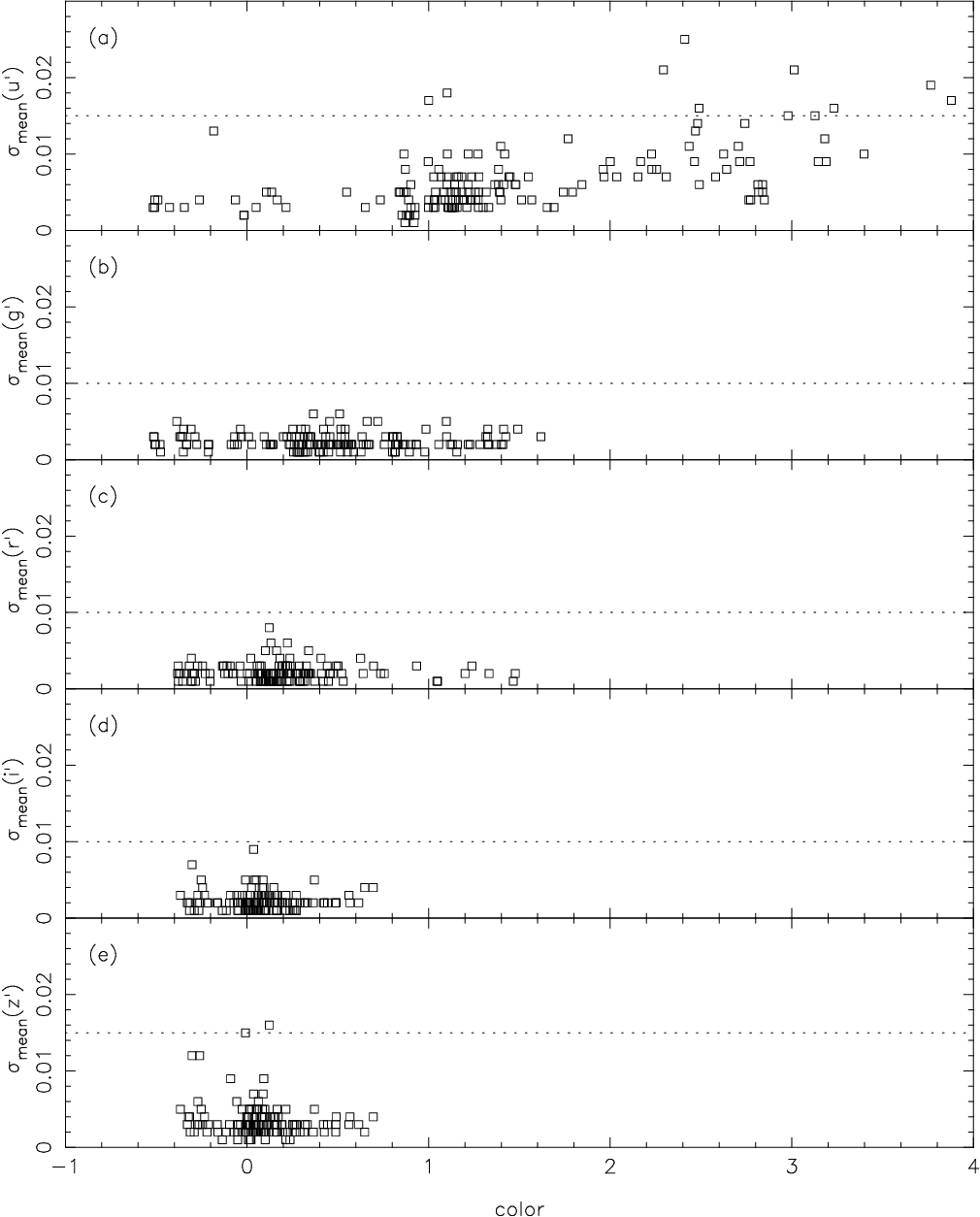}
\caption{Mean error vs.\ color plot 
($\sigma_{\rm mean} (u')$ vs.\ $(u'-g')$, $\sigma_{\rm mean} (g')$
vs.\ $(g'-r')$, $\sigma_{\rm mean} (r')$ vs.\ $(r'-i')$, $\sigma_{\rm
mean} (i')$ vs.\ $(i'-z')$, $\sigma_{\rm mean} (z')$ vs.\ $(i'-z')$).  
The horizontal dashed lines are the
survey requirement for the standard network to meet.
\label{meanerror_vs_color}}
\end{figure}

The above tests --- those showing the residuals from individual
observations plotted against a variety of variables --- were useful in
showing that there are no obvious systematic trends in the
$u'g'r'i'z'$ primary standard star network.  To examine the
system-wide rms errors in the network, we have plotted the standard
error of the mean magnitude for each star in
Figures~\ref{meanerror_vs_mag} and \ref{meanerror_vs_color}.  We have
also listed the minimum, the average, and the maximum of these ``per
star'' errors in Table~\ref{system_errors}, along with the goals
needed to meet the survey end-to-end requirements.
Figure~\ref{meanerror_vs_mag} shows the mean error as a function of
magnitude.  We clearly see in this plot that the fainter standards in
each multiple star field, generally the equatorial fields, have higher
errors.  Figure~\ref{meanerror_vs_color} shows the mean error as a
function of color. Again, we see an increase in the error for the
fainter stars in each multiple star field only this time the errors
increase for the red stars in the $u'$ frames and the blue stars in
the $z'$ frames.  In both of these plots, the horizontal dashed lines
indicate the survey error budget allotted to the standard star network
(the survey goals from Table~\ref{system_errors}).  Note, that, for
all but a handful of stars in $u'$ and $z'$, we more than satisfy the
survey goals.

\placefigure{meanerror_vs_mag}

\placefigure{meanerror_vs_color}

\placetable{system_errors}

\begin{figure}
\figurenum{9}
\epsscale{0.75}
\includegraphics[angle=-90,scale=.75]{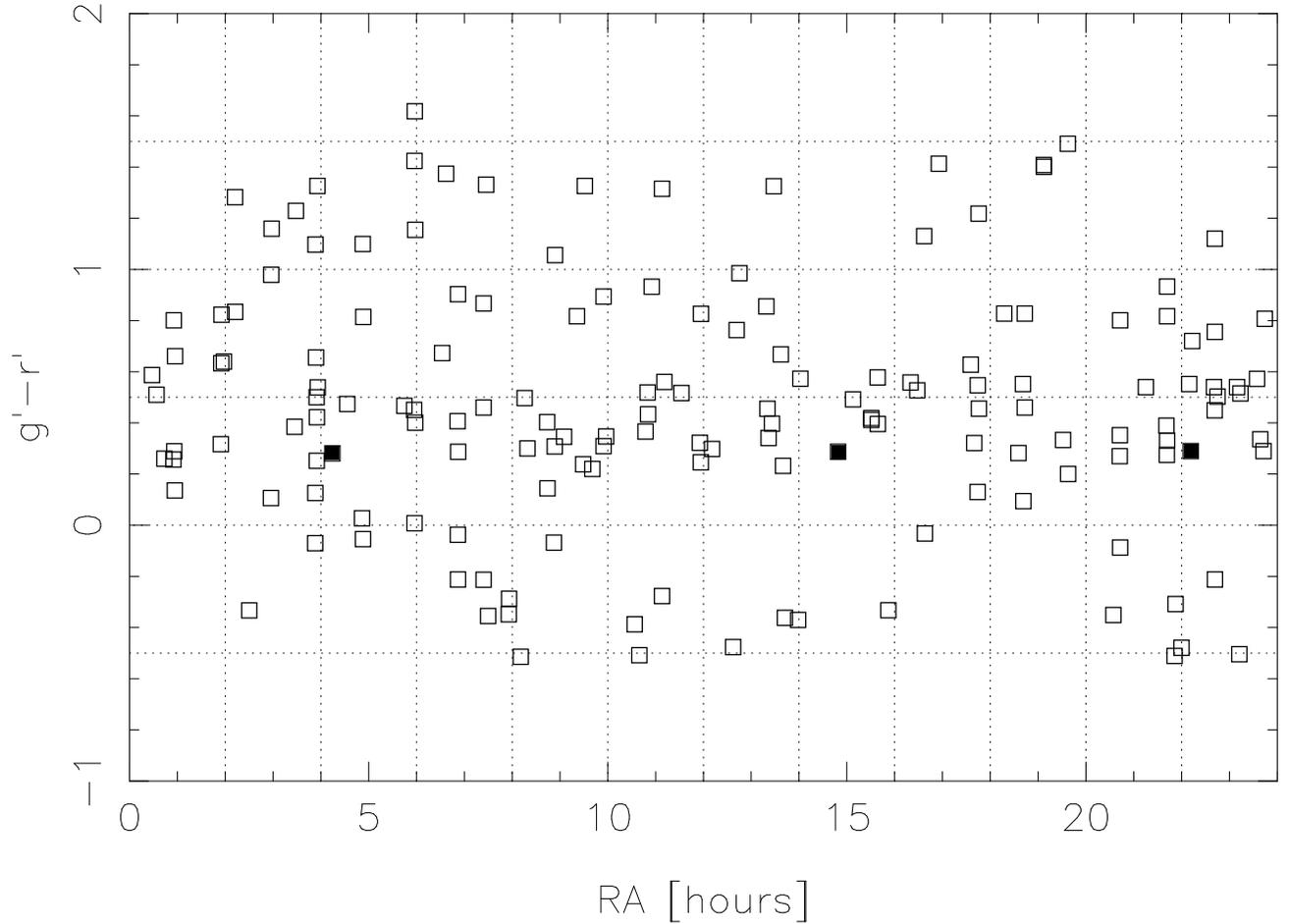}
\caption{The distribution in $g'-r'$ vs.\
right ascension space for the 158 $u'g'r'i'z'$ primary and fundamental
standards (the three filled symbols mark the positions of the three
fundamental standards).  Gridlines demark boxes 0.5~mag wide in
$g'-r'$ and 2~hours wide in right ascension.  Note that almost each box 
in the range $-0.5 < g'-r' < 1.5$ contains one or more standard stars.
\label{color_vs_RA}}
\end{figure}

The distribution of the stars in RA--($g'-r'$) color space is shown in 
Figure~\ref{color_vs_RA}.  While this plot was generally used as an observation 
planning tool during system development, it does show the generally uniform 
distribution of the standards around the sky, at least in the northern 
hemisphere.

\placefigure{color_vs_RA}

\begin{figure}
\figurenum{10}
\epsscale{0.75}
\includegraphics[angle=-90,scale=.75]{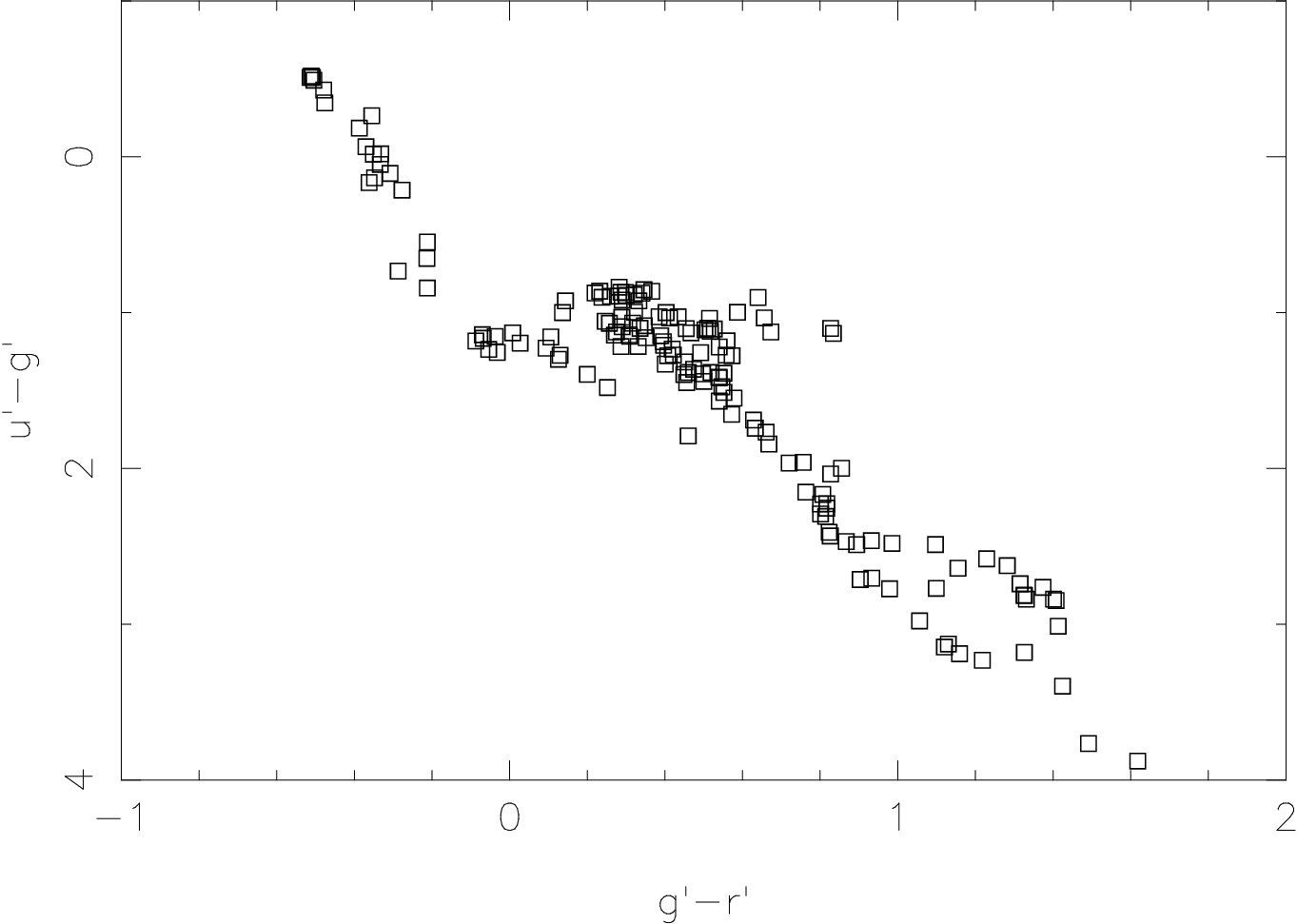}
\caption{The ($u'-g'$) vs.\ ($g'-r'$) color-color 
plot for the 158 $u'g'r'i'z'$ standard stars. 
\label{ugr}}
\end{figure}

\begin{figure}
\figurenum{11}
\epsscale{0.75}
\includegraphics[angle=-90,scale=.75]{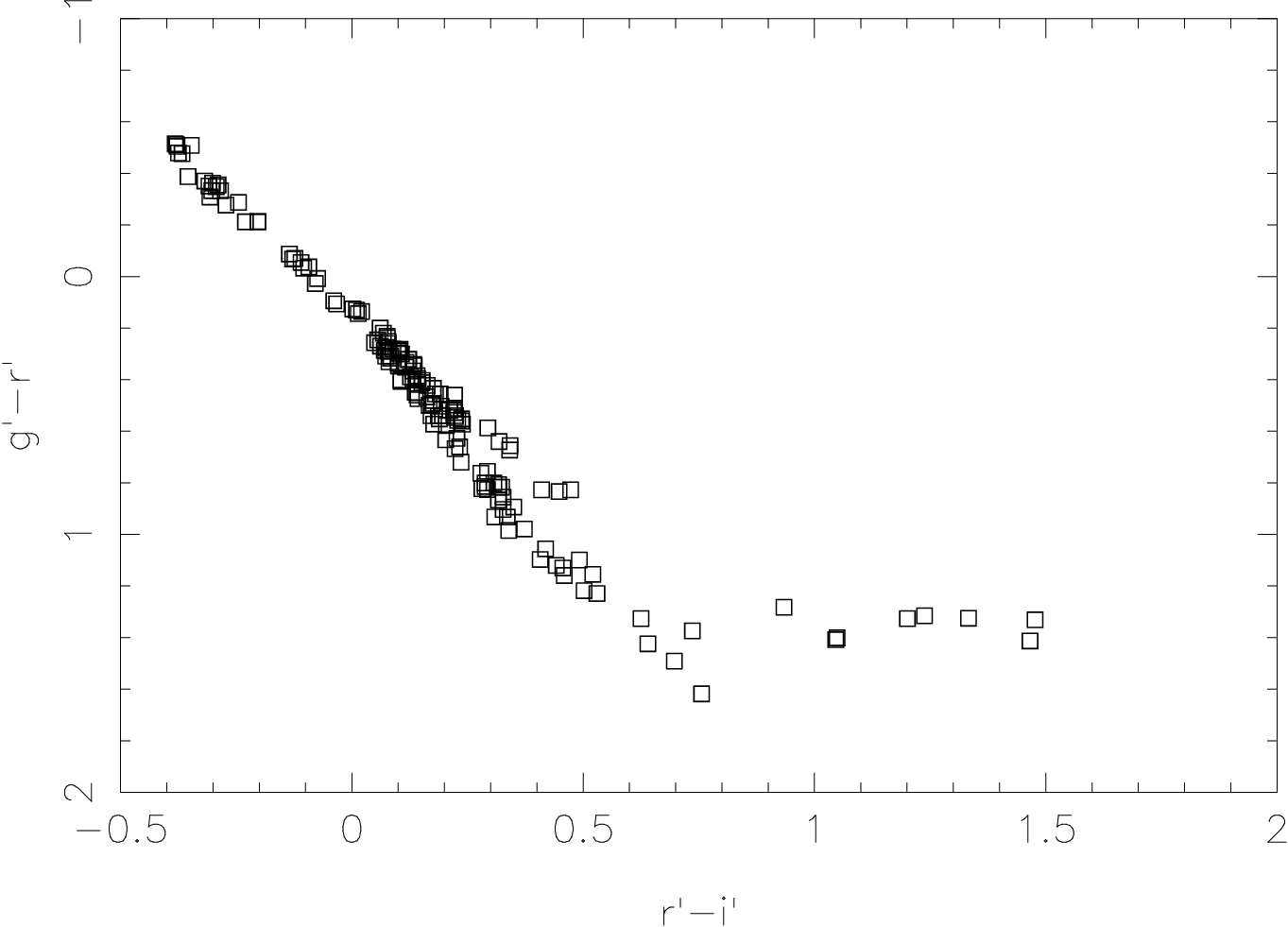}
\caption{The ($g'-r'$) vs.\ ($r'-i'$) color-color 
plot for the 158 $u'g'r'i'z'$ standard stars.
\label{gri}}
\end{figure}

\begin{figure}
\figurenum{12}
\epsscale{0.75}
\includegraphics[angle=-90,scale=.75]{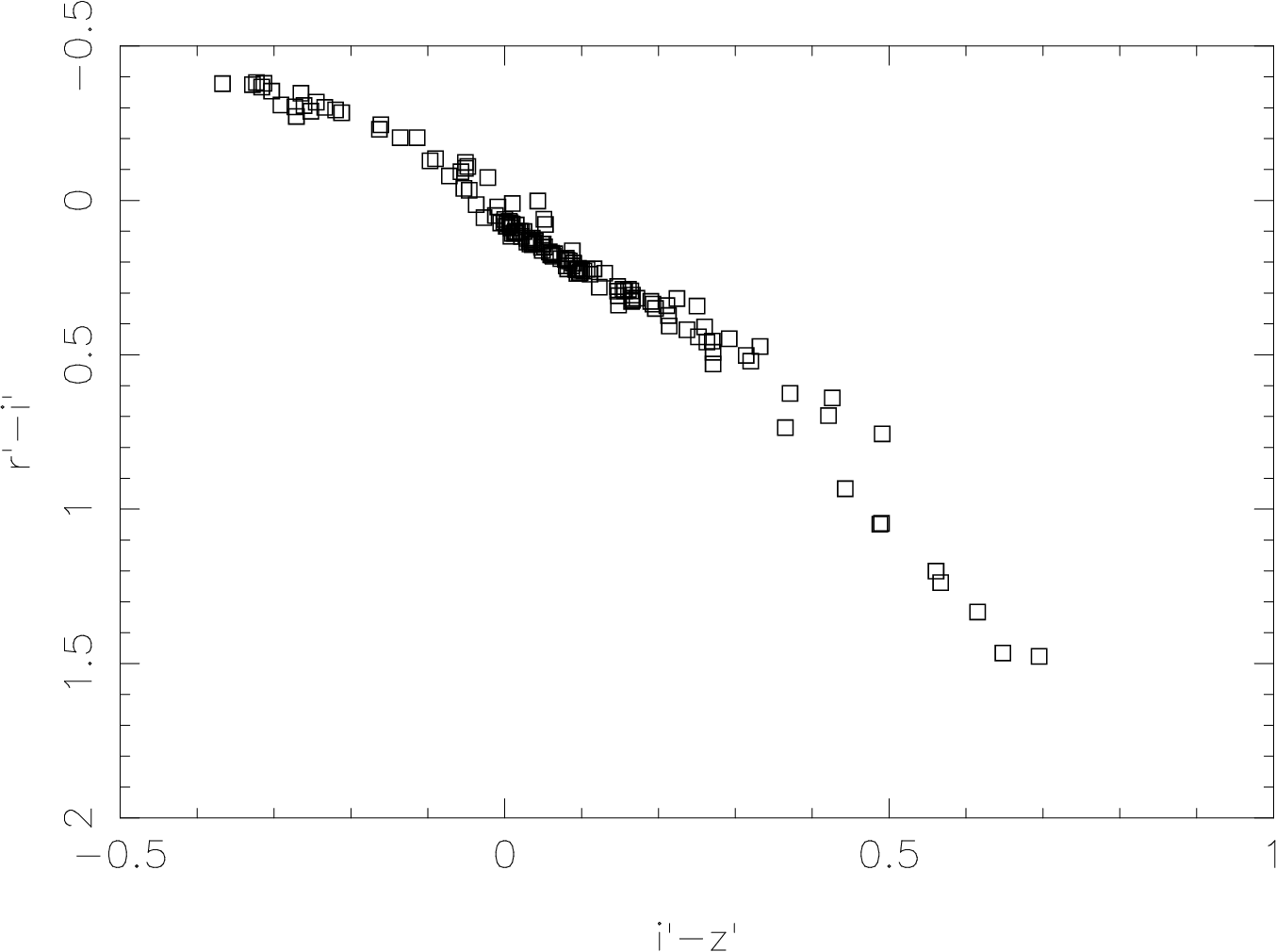}
\caption{The ($r'-i'$) vs.\ ($i'-z'$)
color-color plot for the 158 $u'g'r'i'z'$ standard stars.
\label{riz}}
\end{figure}

The color-color space distributions of the final set of stars are shown in
Figure~\ref{ugr} ($u'-g'$)-vs.-($g'-r'$);
Figure~\ref{gri} ($g'-r'$)-vs.-($r'-i'$); and
Figure~\ref{riz} ($r'-i'$)-vs.-($i'-z'$).  
The break in the linear color transforms at about spectral type M0 is
clearly seen in Figure~\ref{gri} and is also evident in Figure~\ref{riz}.
Separation of the metal-poor stars from the main sequence dwarfs is
seen in Figure~\ref{ugr}.  
The clump of stars in the blue-blue
corner of all three of these plots are the warm-hot white dwarfs.

\placefigure{ugr}
\placefigure{gri}
\placefigure{riz}

\begin{figure}
\figurenum{13}
\epsscale{0.75}
\plotone{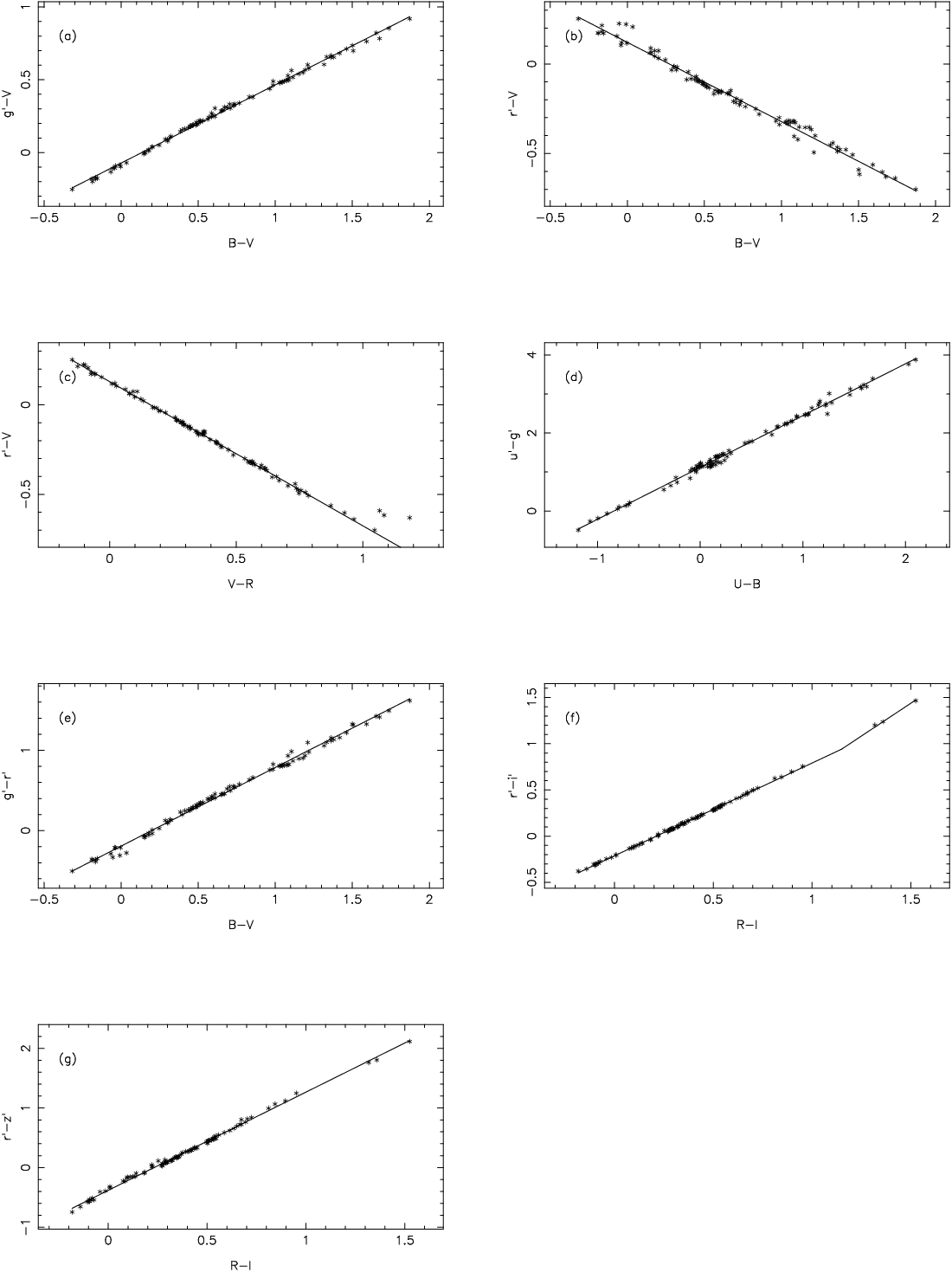}
\caption{Comparison of $u'g'r'i'z'$ and $UBVR_{\rm c}I_{\rm c}$
magnitudes for those $u'g'r'i'z'$ standards measured by Landolt.  The
solid lines denote the linear fits listed under the ``Observed''
column of Table~\ref{UBVRItransform} for the $UBVR_{\rm c}I_{\rm c}
\longrightarrow u'g'r'i'z'$ transformations.
\label{UBVRItransformFigure}}
\end{figure}

\begin{figure}
\figurenum{14}
\epsscale{0.75}
\plotone{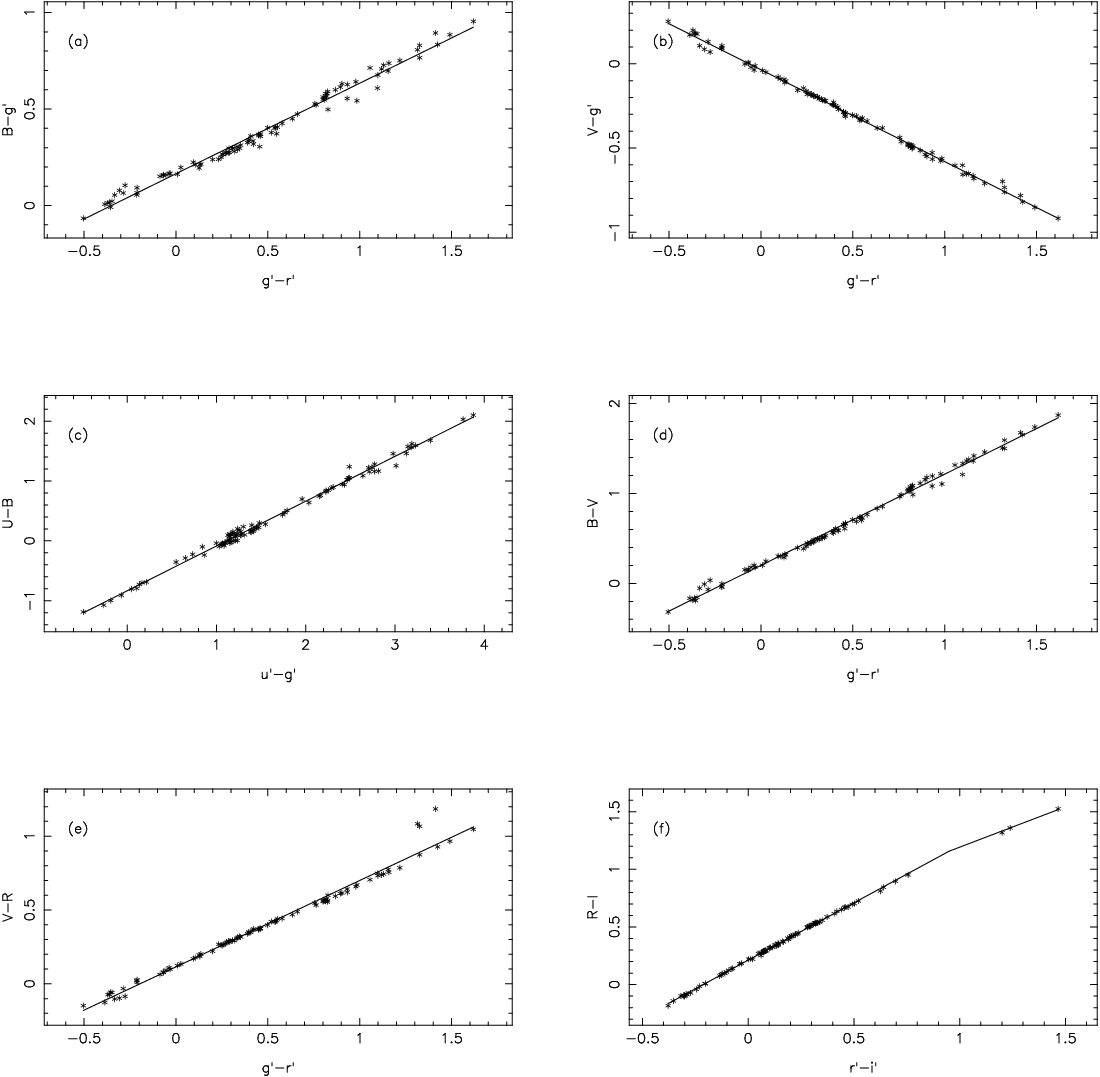}
\caption{Comparison of $u'g'r'i'z'$ and $UBVR_{\rm c}I_{\rm c}$
magnitudes for those $u'g'r'i'z'$ standards measured by Landolt.  The
solid lines denote the linear fits listed under the ``Observed''
column of Table~\ref{UBVRItransform} for the $u'g'r'i'z'
\longrightarrow UBVR_{\rm c}I_{\rm c}$ transformations.
\label{InverseUBVRItransformFigure}}
\end{figure}

Approximate relations for transforming magnitudes and colors
from the Johnson--Morgan--Cousins $UBVR_{\rm c}I_{\rm c}$ system to
the SDSS $u'g'r'i'z'$ system were given in the system defining paper
\citep{fuk96}.  Full details of the development of the relations
presented herein are given by \citet{jorg02}.
\citet{fuk96}'s synthetic transformations and our observed relationships are
given in Table~\ref{UBVRItransform}.  Our observed transformation
relationships are also shown in graphical form in
Figure~\ref{UBVRItransformFigure}, and are similar to those of
\citet{fuk96} (their Figure~6).  We also present relations for
the inverse transformations --- from $u'g'r'i'z'$ to $UBVR_{\rm
c}I_{\rm c}$ --- in Table~\ref{UBVRItransform} and in 
Figure~\ref{InverseUBVRItransformFigure}.

\placetable{UBVRItransform}
\placefigure{UBVRItransformFigure}
\placefigure{InverseUBVRItransformFigure}

\section{The Standard Star Network}

\begin{figure}
\figurenum{15}
\epsscale{0.75}
\plotone{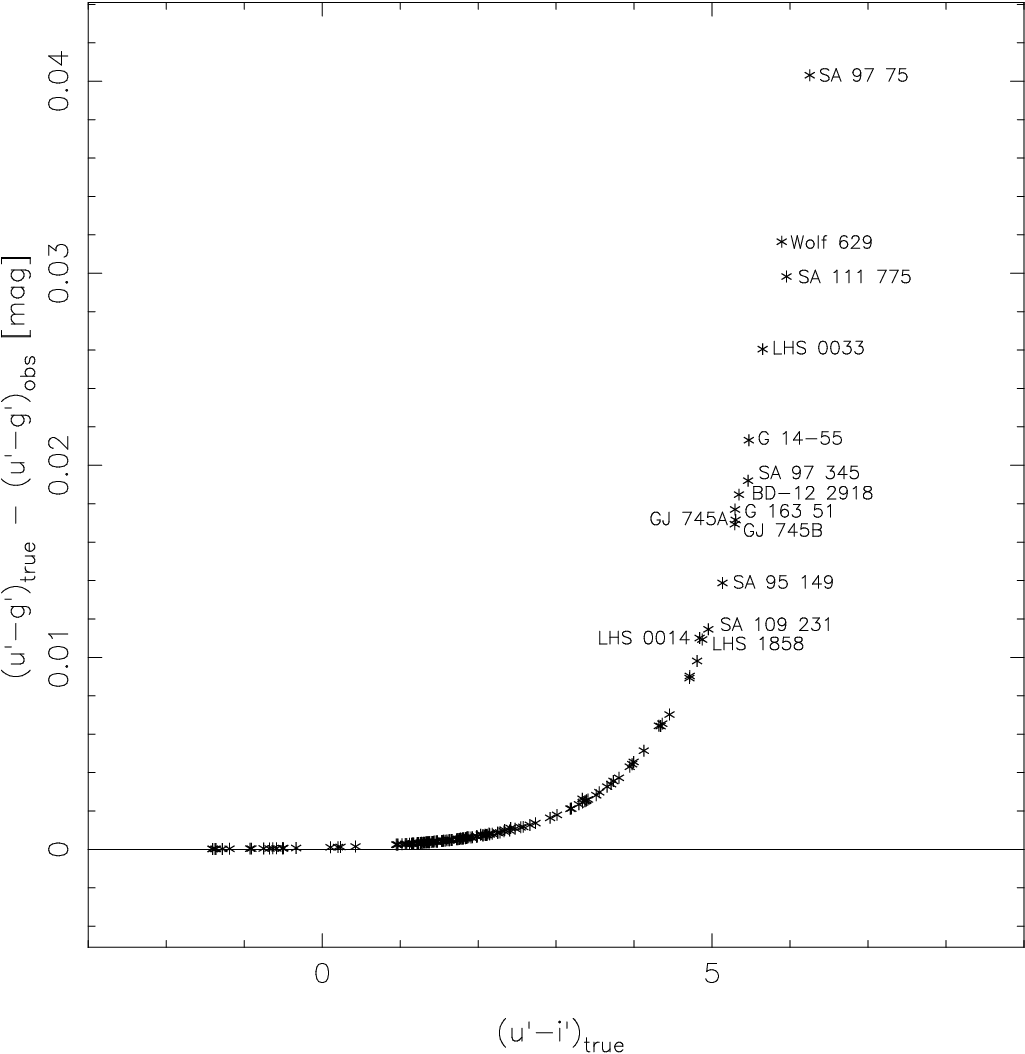}
\caption{The red leak correction [$(u'-g')_{\rm true}-(u'-g')_{\rm obs}$]
vs.\ the red-leak-corrected $(u'-i')$ colors for all 158 $u'g'r'i'z'$ standards.
Note that, for all but a handful of stars, the red leak correction is less
0.01~mag.  Those stars with a red leak correction greater than 0.01~mag are 
explicitly labelled.
\label{redleak}}
\end{figure}

Finally, we present the magnitude and color data for each star along
with astrometric, proper motion and spectroscopic information.  
Table~8 is arranged in order of increasing right ascension and
contains the star name, RA and declination (J2000) in the first three
columns.  The next five columns give the $r'$ magnitudes and the four
color indices.  (N.B.: the $u'-g'$ colors listed here have had the
$u'$ red leak correction of eq.~\ref{redleakcorr2} applied; see
Fig.~\ref{redleak}.)  These five columns are linked with the next five
columns (9-13) which give the standard deviation of the measurements.
As a note, during the reductions we calculated the five filter
magnitudes.  We report colors here as an observational aid.  The
associated uncertainties for the colors are derived from the magnitude
errors added in quadrature.  As such, they may be slightly
overestimated.  The last five columns of this table give the number of
individual measurements, by filter, that were used to determine the
final magnitudes.

\placetable{bigauldtable}

Table~9 is arranged by increasing RA and the
first three columns are the same as Table~8.  Column
4 gives the Guide Star or Hipparcos catalog number, or indicates it is
in the Tycho database.  The coordinate epoch follows in column 5.
Columns 6--10 give the proper motions and uncertainties in
milli-arcsec~yr$^{-1}$ and the reference where these were obtained.  Columns
11 and 12 give the spectral type (where known) and reference.  The
last column indicates that additional notes on a star may be found
in the footnotes to the table.

\placetable{bigauldtabledeux}

\section{Future Work}

Though the setup of the initial primary standard stars for the $u'g'r'i'z'$ 
system is now complete, there is still a large amount of work remaining to 
make this system widely useful to the astronomical community.  Two efforts 
which will continue through the life of the SDSS are reducing the errors in
the mean magnitudes of each standard star and obtaining good magnitudes for
all of the additional stars in each of the standard star fields.  These will
be done by making use of the observations of the SDSS photometric monitoring 
telescope at Apache Point Observatory.  This telescope operates every night
during survey operations to obtain extinction data for survey calibration 
and to transfer the standard star system, through fainter stars, to the 2.5-m
survey data.

In addition to the continued refinement of the primary standard star network, 
two additional areas which require more work are extending the system to 
redder stars and to the southern hemisphere.  The initial standard 
system was limited to stars generally bluer than about dM0, so the redder 
stars are needed to obtain accurate magnitudes for the late M dwarfs and the
new spectral classes, L and T.  An extension to redder stars will make the 
logical tie into the 2MASS survey for the redder objects more meaningful.  
We are developing a program to make this extension,
specifically to look for additional ``knees'' in the color--color
diagrams.  The second area of expansion is into the southern hemisphere.
We began this effort in September 2000 using the 0.9-m at CTIO and the
same observers and reduction software as used in the setup of the
network described in this present paper\footnote{{\tt
http://home.fnal.gov/$\sim$dtucker/Southern\_ugriz/index.html}}.
We undertook this effort to minimize any discrepancies such as those which
have crept into the existing $UBVR_{\rm c}I_{\rm c}$ standards, in
which several independent investigators have been involved in calibrating 
standard stars.

\acknowledgments

The Sloan Digital Sky Survey is a joint project of The University of Chicago, 
Fermilab, the Institute for Advanced Study, the Japan Participation Group, 
The Johns Hopkins University, the Max-Plank-Institute for Astronomy (MPIA), 
the Max-Plank-Institute for Astrophysics (MPA), the New Mexico State 
University, Princeton University, the United States Naval Observatory, and 
the University of Washington.  Apache Point Observatory, site of the SDSS
telescopes, is operated by the Astrophysicsal Research Consortium (ARC).  
\\
\\
Funding for the project has been provided by the Alfred P. Sloan Foundation, 
the SDSS member institutions, the National Aeronautics and Space
Administration, the National Science Foundation, the U.S. Department of 
Energy, the Japanese Monbukagakusho, and the Max Planck Society.
The official SDSS Web site is {\tt http://www.sdss.org}.
\\
\\
The authors would like to thank the USNO-Flagstaff for making the
telescope and detector available for the observations; and Conard
Dahn, Joe Hobart, Fred Harris and Hugh Harris for their assistance
during the observing sessions. We also thank David Hogg, Michael
Strauss, and Gillian Knapp for their valuable reading of the
manuscript and comments.  JAS also acknowledges the many useful
discussions of the project with Conard Dahn and Hugh Harris at USNO,
Arlo Landolt at LSU, and Brian Skiff at Lowell Observatory.  JAS
further acknowledges Lowell Observatory for its support and
hospitality during the observing runs.  JAS and DLT acknowledge
Jeanne Odermann and Sahar Allam for their help in proofreading the text.
Finally, we acknowledge the referee, B. Oke, for his many insightful 
suggestions which have substaintially improved the quality of the paper 
and presentation of the data.
\\
\\
This research has made use of the SIMBAD database, operated at CDS,
Strasbourg, France.

\newpage

\begin{deluxetable}{llrrr}
\tabletypesize{\scriptsize}
\tablecaption{Observing Runs at USNO 1.0-m Telescope\label{obsruns}}
\tablehead{
  \colhead{YYMM} & 
  \colhead{UT Dates} & 
  \colhead{Nights} & 
  \colhead{Clear\tablenotemark{a}} & 
  \colhead{\# Obs.} 
} 
\startdata
9803   & 03/11--03/16 &   6 &  2 &   91  \nl
9804   & 04/10--04/17 &   8 &  2 &   78  \nl
9805   & 05/09--05/18 &  10 &  5 &  213  \nl
9806   & 06/06--06/14 &   9 &  6 &  236  \nl
9807   & 07/07--07/13 &   7 &  1 &   45  \nl
9808   & 08/05--08/11 &   7 &  0 &    0  \nl
9809   & 09/02--09/10 &   9 &  0 &    0  \nl
9810   & 10/02--10/08 &   7 &  3 &  126  \nl
9811   & 10/29--11/09 &  11 &  5 &  195  \nl
9812   & 11/28--12/04 &   7 &  3 &   70  \nl
9901   & 12/31--01/05 &   6 &  3 &  140  \nl
9902   & 01/27--02/04 &   9 &  6 &  212  \nl
9903   & 02/27--03/05 &   7 &  2 &   92  \nl
9904   & 03/27--04/04 &   9 &  3 &  130  \nl
9905   & 04/26--05/03 &   8 &  3 &   58  \nl
9905   & 05/27--05/31 &   5 &  0 &    0  \nl
9906   & 06/26--07/02 &   7 &  5 &  183  \nl
9907   & 07/23--07/28 &   6 &  0 &    0  \nl
9908   & 08/24--08/30 &   7 &  0 &    0  \nl
9909   & 09/22--09/28 &   7 &  4 &  228  \nl
9910   & 10/20--10/28 &   9 &  6 &  215  \nl
9911   & 11/18--11/25 &   8 &  2 &   97  \nl
9912   & 12/17--12/23 &   7 &  3 &  116  \nl
0001   & 01/21--01/27 &   7 &  0 &    0  \nl
\hline
$\sum$ &              & 183 & 64 & 2525  \nl
\enddata

\tablenotetext{a}{Here, a ``clear'' night is as judged at the telescope.}
\end{deluxetable}

\newpage

\begin{deluxetable}{lllrrrrr} 
\tablecaption{The $u'g'r'i'z'$ Fundamental Standards \label{fundamentals}}
\tablehead{
  \colhead{Star\tablenotemark{a}}&
  \colhead{R.A.} & 
  \colhead{Dec.} & 
  \colhead{$u'$} & 
  \colhead{$g'$} & 
  \colhead{$r'$} & 
  \colhead{$i'$} & 
  \colhead{$z'$} 
}
\startdata
BD+17$\arcdeg$4708&22:11:30 &+18:05:30 &10.56 & 9.64 & 9.35 & 9.25 & 9.23 \nl
\nl
BD+26$\arcdeg$2606 &14:49:02 &+25:42:27 &10.761 & 9.891 & 9.604 & 9.503 & 9.486 \nl
$\sigma$            &         &          & 0.001 & 0.001 & 0.001 & 0.001 & 0.001 \nl
BD+21$\arcdeg$0607 &04:14:35 &+22:21:06 &10.289 & 9.395 & 9.114 & 9.025 & 9.017 \nl
$\sigma$            &         &          & 0.002 & 0.001 & 0.001 & 0.001 & 0.001 \nl
\enddata

\tablenotetext{a} {The magnitudes for BD+17$\arcdeg$4708 are defined as given 
in this table.}
\end{deluxetable}

\newpage

\begin{deluxetable}{rrr} 
\tablecaption{The Cosmic Colors \label{cosmic_colors}}
\tablewidth{0pt}
\tablehead{
  \colhead{color} & 
  \colhead{value} & 
  \colhead{scatter}
}
\startdata
 $(u'-g')$ & 1.42 & $\pm$0.82 \nl
 $(g'-r')$ & 1.11 & $\pm$0.46 \nl
 $(r'-i')$ & 0.48 & $\pm$0.34 \nl
 $(i'-z')$ & 0.35 & $\pm$0.22 \nl
\enddata
\end{deluxetable}

\newpage

\begin{deluxetable}{lcc} 
\tabletypesize{\normalsize}
\tablecaption{Second Order Extinction Coefficients \label{cterms}}
\tablewidth{0pt}
\tablehead{
  \colhead{Filter} & 
  \colhead{superExcal} & 
  \colhead{solve\_network}
}
\startdata
$c_{u'}$ & $-0.021\pm0.003$ & $-0.032$ \nl
$c_{g'}$ & $-0.016\pm0.003$ & $-0.015$ \nl
$c_{r'}$ & $-0.004\pm0.003$ & $+0.000$ \nl
$c_{i'}$ & $+0.006\pm0.003$ & $+0.005$ \nl
$c_{z'}$ & $+0.003\pm0.003$ & $+0.006$ \nl
\enddata
\end{deluxetable}

\newpage

\begin{deluxetable}{lcccc} 
\tabletypesize{\normalsize}
\tablecaption{Mean Errors in Calibrated Magnitudes\label{system_errors}}
\tablewidth{0pt}
\tablehead{
  \colhead{Filter} & 
  \colhead{Survey Goal} & 
  \multicolumn{3}{c}{Calculated} \\
  &
  & 
  \colhead{minimum} &
  \colhead{average} &
  \colhead{maximum}
}
\startdata
$\sigma_{u'}$ & 0.015 & 0.001 & 0.007 & 0.025 \nl
$\sigma_{g'}$ & 0.010 & 0.001 & 0.002 & 0.006 \nl
$\sigma_{r'}$ & 0.010 & 0.001 & 0.002 & 0.008 \nl
$\sigma_{i'}$ & 0.010 & 0.001 & 0.002 & 0.009 \nl
$\sigma_{z'}$ & 0.015 & 0.001 & 0.003 & 0.016 \nl
\enddata
\end{deluxetable}

\newpage

\begin{deluxetable}{ccc} 
\tabletypesize{\footnotesize}
\tablecaption{Transformations Between $UBVR_{\rm c}I_{\rm c}$ and $u'g'r'i'z'$ \label{UBVRItransform}}
\tablehead{
  \colhead{Mag/Color} & 
  \colhead{Observed} & 
  \colhead{Synthetic\tablenotemark{a}} 
}
\startdata
\multicolumn{3}{c}{$UBVR_{\rm c}I_{\rm c} \longrightarrow u'g'r'i'z'$} \nl
        &                       &                       \nl
$g'$ & = $V + 0.54(B-V) - 0.07 $ & = $V + 0.56(B-V) - 0.12 $ \nl
$r'$ & = $V - 0.44(B-V) + 0.12 $ & = $V - 0.49(B-V) + 0.11 $ \nl
$r'$ for $(V-R) < 1.00$   & = $V - 0.81(V-R) + 0.13 $ & = $V - 0.84(V-R) + 0.13 $ \nl
$r'$ for $(V-R) \ge 1.00$ &           \nodata         & = $V - 0.84(V-R) + 0.13 $ \nl
$u'-g'$ & = $1.33(U-B) + 1.12 $ & = $1.38(U-B) + 1.14 $ \nl
$g'-r'$ & = $0.98(B-V) - 0.19 $ & = $1.05(B-V) - 0.23 $ \nl
$r'-i'$ for $(R-I) < 1.15$  & = $1.00(R-I) - 0.21 $ & = $0.98(R-I) - 0.23 $ \nl
$r'-i'$ for $(R-I)\ge 1.15$ & = $1.42(R-I) - 0.69 $ & = $1.40(R-I) - 0.72 $ \nl
$r'-z'$ for $(R-I) < 1.65$  & = $1.65(R-I) - 0.38 $ & = $1.59(R-I) - 0.40 $ \nl
$r'-z'$ for $(R-I)\ge 1.65$ & none observed         & = $2.64(R-I) - 2.16 $ \nl
        &                       &                       \nl
        &                       &                       \nl
\multicolumn{3}{c}{$u'g'r'i'z' \longrightarrow UBVR_{\rm c}I_{\rm c}$} \nl
        &                       &                       \nl
$B$ & = $g' + 0.47(g'-r') + 0.17$ & \nodata \nl
$V$ & = $g' - 0.55(g'-r') - 0.03$ & \nodata \nl
$U-B$ & = $0.75(u'-g') - 0.83$    & \nodata \nl 
$B-V$ & = $1.02(g'-r') + 0.20$    & \nodata \nl
$V-R$ & = $0.59(g'-r') + 0.11$    & \nodata \nl
$R-I$ for $(r'-i') <   0.95$  & = $1.00(r'-i') + 0.21$  &  \nodata\nl
$R-I$ for $(r'-i') \ge 0.95$  & = $0.70(r'-i') + 0.49$  &  \nodata\nl
        &                       &                       \nl
\enddata
\tablenotetext{a}{From \citet{fuk96}.}
\end{deluxetable}

\newpage

\begin{deluxetable}{lr@{ }rrrrrrr@{ }r@{ }r@{ }r@{ }rr@{ }r@{ }r@{ }r@{ }r}
\rotate
\tabletypesize{\tiny}
\tablecaption{The $u'g'r'i'z'$ standard star network: calibrated magnitudes and colors. \label{bigauldtable}}
\tablewidth{0pt}
\tablenum{8}
\tablehead{
  \colhead{Star Name}  & 
  \colhead{RA (J2000.0)}  & 
  \colhead{DEC (J2000.0)}  &
  \colhead{$r'$}  & 
  \colhead{$u'-g'$}  & 
  \colhead{$g'-r'$}  & 
  \colhead{$r'-i'$}  & 
  \colhead{$i'-z'$}  & 
  \colhead{$\sigma_{r'}$}  & 
  \colhead{$\sigma_{u'-g'}$}  & 
  \colhead{$\sigma_{g'-r'}$}  & 
  \colhead{$\sigma_{r'-i'}$}  & 
  \colhead{$\sigma_{i'-z'}$}  & 
  \colhead{$n_{u'}$}  & 
  \colhead{$n_{g'}$}  & 
  \colhead{$n_{r'}$}  & 
  \colhead{$n_{i'}$}  & 
  \colhead{$n_{z'}$} 
}
\startdata
     Hilt~31 &   00:28:11.15  & $+$64:07:51.8  &      10.996 &    0.999 &    0.587 &    0.294 &    0.147 &       0.001 &    0.003 &    0.001 &    0.001 &    0.002 &        16 &    16 &    16 &    15 &    15 \\ 
   G~158-100 &   00:33:54.60  & $-$12:07:58.9  &      14.691 &    1.101 &    0.510 &    0.222 &    0.092 &       0.006 &    0.019 &    0.008 &    0.007 &    0.010 &         6 &     7 &     8 &     6 &     6 \\ 
    BD+71~0031 &   00:43:44.34  & $+$72:10:43.1  &      10.091 &    0.896 &    0.260 &    0.079 &    0.007 &       0.001 &    0.003 &    0.002 &    0.002 &    0.004 &        14 &    14 &    14 &    14 &    14 \\ 
   SA~92~342 &   00:55:09.90  & $+$00:43:12.9  &      11.527 &    1.069 &    0.257 &    0.049 & $-$0.012 &       0.001 &    0.004 &    0.001 &    0.001 &    0.002 &        28 &    27 &    27 &    26 &    25 \\ 
   SA~92~263 &   00:55:39.41  & $+$00:36:20.0  &      11.467 &    2.229 &    0.801 &    0.307 &    0.166 &       0.001 &    0.008 &    0.001 &    0.002 &    0.003 &        23 &    27 &    27 &    26 &    26 \\ 
   SA~92~502 &   00:56:08.13  & $+$01:04:25.1  &      11.712 &    1.031 &    0.289 &    0.078 &    0.010 &       0.002 &    0.005 &    0.004 &    0.003 &    0.004 &         9 &     9 &     9 &     9 &     9 \\ 
   SA~92~282 &   00:56:46.86  & $+$00:38:30.9  &      12.936 &    1.000 &    0.136 &    0.021 & $-$0.009 &       0.004 &    0.017 &    0.004 &    0.006 &    0.016 &         5 &     6 &     5 &     5 &     3 \\ 
   SA~92~288 &   00:57:17.00  & $+$00:36:48.7  &      11.350 &    1.768 &    0.661 &    0.233 &    0.098 &       0.002 &    0.013 &    0.005 &    0.003 &    0.005 &         6 &     6 &     6 &     5 &     5 \\
\enddata

\end{deluxetable}

\newpage

\vspace{5.0cm}
\begin{deluxetable}{l@{ }r@{ }r@{ }r@{ }r@{ }r@{ }r@{ }r@{ }r@{ }r@{ }r@{ }r@{ }c@{ }}
\rotate
\tabletypesize{\tiny}
\tablecaption{The $u'g'r'i'z'$ standard star network: astrometry, spectral properties, and notes. \label{bigauldtabledeux}}
\tablewidth{0pt}
\tablenum{9}
\tablehead{
  \colhead{Star Name}             & 
  \colhead{$\alpha$ (J2000.0)}    & 
  \colhead{$\delta$ (J2000.0)}    &
  \colhead{T-Tycho/GSC/Hipp}      & 
  \colhead{Epoch}                 & 
  \colhead{$\mu(\alpha \cos \delta)$}            & 
  \colhead{$\sigma(\mu(\alpha \cos \delta))$}    & 
  \colhead{$\mu(\delta)$}         & 
  \colhead{$\sigma(\mu(\delta))$} & 
  \colhead{pm}                    & 
  \colhead{spectral}              & 
  \colhead{spectrum}              &
  \colhead{Notes}                 \\
  \colhead{}                      & 
  \colhead{}                      & 
  \colhead{}                      &
  \colhead{\#}                    & 
  \colhead{year}                  & 
  \colhead{marcsec/yr}            & 
  \colhead{marcsec/yr}            & 
  \colhead{marcsec/yr}            & 
  \colhead{marcsec/yr}            & 
  \colhead{ref.}                  & 
  \colhead{type}                  & 
  \colhead{ref.}                  &
  \colhead{}                      \\
}
\startdata
     Hilt~31 &   00:28:11.15  & $+$64:07:51.8  &           T & 2000.000 &   $-$3.90 &  3.40 &   $-$4.90 &  3.50 &  TYC2 &      B1V &   H56  &  1 \\
   G~158-100 &   00:33:54.60  & $-$12:07:58.9  & G0527100531 & 2000.000 &   152.00 &  3.00 & $-$179.00 &  4.00 &  HD80 &       DG &   G84  &  w,2 \\
    BD+71~0031 &   00:43:44.34  & $+$72:10:43.1  &        3430 & 2000.000 &   321.50 &  1.80 &    90.70 &  1.80 &  TYC2 &     sdF5 &   C83  &  m \\
   SA~92~342 &   00:55:09.90  & $+$00:43:12.9  &           T & 2000.000 &   $-$2.20 &  1.70 &   $-$0.20 &  1.70 &  TYC2 &       F5 &  DL79  &  \\
   SA~92~263 &   00:55:39.41  & $+$00:36:20.0  & G0001200675 & 1983.702 &   \nodata&\nodata&   \nodata&\nodata&   N/A &    G8III &  DL79  &  \\
   SA~92~502 &   00:56:08.13  & $+$01:04:25.1  &           T & 2000.000 &     7.20 &  1.70 &     7.40 &  1.70 &  TYC2 &      N/A &   N/A  &  \\
   SA~92~282 &   00:56:46.86  & $+$00:38:30.9  &           T & 2000.000 &    28.30 &  2.80 &     6.10 &  2.70 &  TYC2 &       A5 &  DL79  &  \\
   SA~92~288 &   00:57:17.00  & $+$00:36:48.7  &           T & 2000.000 &     4.90 &  1.40 &     4.90 &  1.40 &  TYC2 &      K3V &  DL79  &  \\
\enddata
\end{deluxetable}
\clearpage
\begin{tiny}
\begin{tabular}{l}
$^{\rm m}$ High proper motion star.\\
$^{\rm p}$ HST \&/or IUE photometric standard.\\
$^{\rm s}$ HST \&/or FUV spectrophotometric standard.\\
$^{\rm w}$ White dwarf star.\\
\\
$^{\rm 1}$ Heavily reddened star.\\
$^{\rm 2}$ Not listed in Villanova WD catalog (MS99); =EGGR-382, LTT-300, G270-8, USNO-490.\\
$^{\rm 3}$ Blue straggler in Per OB1 association.\\
\end{tabular}
\end{tiny}

\newpage

\clearpage

\clearpage

\clearpage

\clearpage

\clearpage

\clearpage

\clearpage

\clearpage

\clearpage

\clearpage

\clearpage

\clearpage

\clearpage

\clearpage

\clearpage


\newpage

\begin{thebibliography}{}

\bibitem[Bemporad(1904)]{bem04} Bemporad, A. 1904, Mitteil. Grossherzogl.
    Sternwarte, Heidelberg, 4 

\bibitem[Berger \& Fringant(1980)]{BH80} Berger, J., \& Fringant, A. M. 
    1980, \aap, 85, 367 (BH80)

\bibitem[Bidelman(1985)]{B85} Bidelman, W. P. 1985, \apjs, 59, 197 (B85)

\bibitem[Cannon \& Pickering(1918--24)]{HD} 
	Cannon, A. J., \& Pickering, E. C. 1918--24, 
	Henry Draper Catalogue and Extension,
	Harv.\ Ann. 91--100  (HD)

\bibitem[Carney(1983)]{C83} Carney,  B. W. 1983, \aj, 88, 610  (C83)

\bibitem[Carney \& Latham(1987)]{CL87} Carney, B. W., \& Latham, D. W. 1987, 
    \aj, 93, 116 (CL87)

\bibitem[Cowley, Hiltner \& Witt(1967)]{CH67} Cowley, A. P., Hiltner, W. A, 
    \& Witt, A. N. 1967, \aj, 72, 1334  (CH67)

\bibitem[Doyle \& Butler(1990)]{DB90} Doyle, J. G., \& Butler, C. J. 1990, 
    \aap, 235, 335 (DB90)

\bibitem[Drilling \& Landolt(1979)]{DL79} Drilling, J. S., \& Landolt, A. U.
    1979, \aj, 84, 783 (DL79)

\bibitem[Duquennoy \& Mayor(1991)]{DM91} Duquennoy, A., \& Mayor, M. 1991, 
    \aap, 248, 485 (DM91)

\bibitem[Eggen \& Sandage(1959)]{ES59} Eggen, O. J., \& Sandage, A. 1959, 
    \mnras, 119, 255 (ES59)

\bibitem[Elkin(1996)]{E96} Elkin, V. G. 1996, \aap, 312L, 5 (E96)

\bibitem[Fukugita et~al.(1996)]{fuk96} Fukugita, M., Ichikawa, T., 
    Gunn, J. E., Doi, M., Shimasaku, K., \& Schneider, D. P. 1996, \aj, 
    111, 1748

\bibitem[Giclas, Burnham \& Thomas(1978)]{G78} Giclas, H. L., 
    Burnham, R., Jr., \& Thomas, N. G. 1978, Lowell Proper Motion Survey, 
    Southern Hemisphere Catalog, Lowell Observatory Bulletin, No. 164,
    8, 89 (G78)

\bibitem[Gizis(1997)]{G97} Gizis, J. E. 1997, \aj, 113, 806 (G97)

\bibitem[Greenstein(1954)]{G54} Greenstein, J. L. 1954, \pasp, 66, 126 (G54)

\bibitem[Greenstein(1984)]{G84} Greenstein, J. L. 1984, \apj, 276, 602 (G84)

\bibitem[Greenstein \& Sargent(1974)]{GS74} Greenstein, J. L., \& 
    Sargent, A. I. 1974, \apjs, 28, 157 (GS74)

\bibitem[Gunn et~al.(1998)]{GCRS98} Gunn, J. E., 
    Carr, M., Rockosi, C., Sekiguchi, M., et~al.  1998, \aj, 116, 3040

\bibitem[Harrington \& Dahn(1980)]{HD80} Harrington, R. S., \&  Dahn, C. C. 
    1980, \aj, 85, 454 (HD80)

\bibitem[Heintz(1993)]{He93} Heintz, W. D. 1993, \aj, 105, 1188 (He93)

\bibitem[Heintz(1994)]{He94} Heintz, W. D. 1994, \aj, 108, 2338 (He94)

\bibitem[Henry, Kirkpatrick \& Simons(1994)]{HK94} Henry, T. J., 
    Kirkpatrick, J. D., \& Simons, D. A. 1994, \aj, 108, 1437 (HK94)

\bibitem[Hiltner(1956)]{H56} Hiltner, W. A. 1956, \apjs, 2, 389 (H56)

\bibitem[Holberg, Bastrow \& Sion(1998)]{HBS98} Holberg, J. B, 
    Bastrow, M. A., \& Sion, E. M. 1998, \apjs, 119, 207 (HBS98)

\bibitem[H{\o}g et al.(2000)]{TYC2} H{\o}g, E., Fabricius, C., 
    Makarov, V. V. et al.\ 2000, \aap, 355, L27 (TYC2)

\bibitem[Houk \& Smith-Moore(1988)]{MSS88} Houk, N., \& Smith-Moore, M. 
    1988, Michigan Catalogue of Two-Dimensional Spectral Types for the
    HD Stars, Vol. 4, Univ. of Michigan, Ann Arbor (MSS88)

\bibitem[Houk \& Swift(1999)]{MSS99} Houk, N., \& Swift, C. 
    1999, Michigan Catalogue of Two-Dimensional Spectral Types for the HD 
    Stars, Vol. 5, Univ. of Michigan, Ann Arbor (MSS99)

\bibitem[Iriarte(1958)]{Ir58} Iriarte, B. 1958, \apj, 127, 507 (Ir58)

\bibitem[Jones(1966)]{J66} Jones, D. H. P. 1966, R. Obs. Bull. No. 126 (J66)

\bibitem[Johnson \& Morgan(1953)]{JM53} Johnson, H. L., \& Morgan, W. W. 1953, 
	\apj, 117, 313 (JM53)

\bibitem[Jorgensen et al.(2002)]{jorg02} Jorgensen, A., et al. 2002, in preparation

\bibitem[Kent(1985)]{kent85} Kent, S. 1985, \pasp, 97, 165

\bibitem[Landolt(1973)]{lan73}  Landolt, A. U. 1973, \aj, 78, 959 (L73)

\bibitem[Landolt(1983)]{lan83}  Landolt, A. U. 1983, \aj, 88, 439

\bibitem[Landolt(1992)]{lan92}  Landolt, A. U. 1992, \aj, 104, 372

\bibitem[Lasker et al.(1996--99)]{GSC} 
	Lasker, B. M., Russel, J. N., \& Jenkner, H. 1996--99, 
	The Guide Star Catalog Version 1.1-ACT (GSC-ACT Catalogue), 
	The Association of Universities for Research in Astronomy, Inc.  (GSC)

\bibitem[Little et~al.(1995)]{LD95}  Little, J. E., Dufton, P. L., 
    Keenan, F. P., Hambly, N. C., Conlon, E. S., Brown, P. J. F., \& 
    Miller, L. 1995, \apj, 447, 783 (LD95)

\bibitem[Luyten(1979)]{NLTT}  Luyten, W. J. 1979, New Luyten Catalogue 
    of Stars with Proper Motions Larger than Two Tenths of an Arcsecond 
    Vol 1 \& 2, Univ. of Minnesota, Minneapolis $+$ Vol. 3 \& 4 (1980) (NLTT)

\bibitem[McCook \& Sion(1987)]{GE87}  McCook, G. P., \& Sion, E. M. 
    1987, \apjs, 65, 603 (GE87)

\bibitem[McCook \& Sion(1999)]{McCS99} McCook, G. P. \& Sion, E. M. 
    1999, \apjs, 121, 1 (MS99)

\bibitem[Mermilliod, Mermilliod, \& Hauck(1997)]{MMH97} 
    Mermilliod, J.-C., Mermilliod, M., \& Hauck, B. 1997, \aaps, 124, 349

\bibitem[Oke(1990)]{oke90} Oke, J. B. 1990, \aj, 99, 1621

\bibitem[Oke \& Gunn(1983)]{OG83} Oke, J. B., \& Gunn, J. E. 1983, \apj, 
    266, 713

\bibitem[Perryman \& ESA(1997)]{HIPTYC1} Perryman, M. A. C. \& 
	ESA 1997, The Hipparcos and Tycho catalogues. Astrometric and
	photometric star catalogues derived from the ESA Hipparcos
	Space Astrometry Mission, Publisher: Noordwijk, Netherlands:
	ESA Publications Division, 1997, Series: ESA SP Series vol no:
	1200, ISBN: 9290923997 (set) (HIP, TYC1)

\bibitem[Popper(1942)]{P42}  Popper, D. 1942, \apj, 95, 307 (P42)

\bibitem[Popper(1943)]{P43}  Popper, D. 1943, \apj, 98, 209 (P43)

\bibitem[Reid, Hawley, \& Gizis(1995)]{MSU} Reid, I. N., Hawley, S. L., 
    \& Gizis, J. E. 1995, \aj, 110, 1838 (MSU)

\bibitem[Roman(1955)]{Ro55}  Roman, N. G. 1955, \apjs, 2, 195 (Ro55)

\bibitem[Sandage(1964)]{sand64} Sandage, A. 1964, \apj, 139, 442

\bibitem[Sandage(1969)]{S69} Sandage, A. 1969, \apj, 158, 1115 (S69)

\bibitem[Sergey et~al.(1996)]{ser96} 
	Sergey, G., Berman, E., Huang, C.-H., Kent, S., Newberg, H., 
	Nicinski, T., Petravick, D., \& Stoughton, C. 1996, in 
    	Astronomical Data Analysis Software and Systems V, 
	ASP Conf.\ Ser., Vol.~101, ed. G.~H. Jacoby \& J. Barnes, 
	(San Francisco:ASP), 248

\bibitem[Smith et al.(2002)]{smith02} Smith, J. A., et al. 2002, in preparation

\bibitem[SAO(1966)]{SAO} Smithsonian Astrophysical Observatory Staff 1966,
	United States Naval Observatory, Astronomical Data Center 1990, 
	SAO Star Catalog J2000 (SAO)

\bibitem[Stone(1977)]{stone77} Stone, R. P. S. 1977, \apj, 218, 767

\bibitem[Stone(1996)]{S96}  Stone, R. P. S. 1996, \apjs, 107, 423 (S96)

\bibitem[Stoughton(1995)]{sto95} Stoughton, C. 1995, \baas, 187, 9101

\bibitem[Stoughton et al.(2001)]{sto01} Stoughton, C., Lupton, R. H., 
	Bernardi, M. et al. 2001, \aj, in press

\bibitem[Thejll(1997)]{TF97}  Thejll, P. 1997, \aap, 317, 689 (TF97)

\bibitem[Thuan \& Gunn(1976)]{TG76} Thuan, T. X., \& Gunn, J. E. 1976, 
    \pasp, 88, 543

\bibitem[Tucker et~al.(2002)]{dlt00} Tucker, D. L., et al. 2002, in preparation

\bibitem[Turnshek et~al.(1990)]{T90}  Turnshek, D. A., Bohlin, R. C., 
    Williamson, R. L., II, Lupie, O. L., \& Koornneef, J. 1990, \aj, 
    99, 1243 (T90)

\bibitem[van~Altena(1995)]{YTP95} van Altena, W., Lee, J. T., \& Hoffleit, D., 
    The General Catalogue of Trigonometric (Stellar) Parallaxes, 
    Fourth Edition, 1995, Yale University, New Haven (YTP95)

\bibitem[Veeder(1974)]{vee74} Veeder, G. J. 1974, \aj, 79, 1056

\bibitem[Weis(1994)]{W94}  Weis, E. W. 1994, \aj, 107, 1135 (W94)

\bibitem[York et~al.(2000)]{york00} York, D. G., Adelman, J., 
	Anderson, J. E. et al.  2000, \aj, 120, 1579

\end{thebibliography}
\end{document}